\documentclass[twocolumn,aps,groupedaddress]{revtex4}

\usepackage{graphicx}
\usepackage{dcolumn}
\usepackage{bm}
\usepackage{amsmath,amssymb,amsfonts}
\usepackage{color}
\usepackage{ulem}

\begin{document}

\title{Unconventional Superconductivity and Density Waves in Twisted Bilayer Graphene}
\author{Hiroki Isobe}
\author{Noah F. Q. Yuan}
\author{Liang Fu}
\affiliation{Department of Physics, Massachusetts Institute of Technology, Cambridge, Massachusetts 02139, USA}

\begin{abstract}
We study electronic ordering instabilities of twisted bilayer graphene with $n=2$ electrons per supercell, where correlated insulator state and superconductivity are recently observed. Motivated by the Fermi surface nesting and the proximity to Van Hove singularity, we introduce a hot-spot model to study the effect of various electron interactions systematically. Using renormalization group method, we find $d$/$p$-wave superconductivity and charge/spin density wave emerge as the two types of leading instabilities driven by Coulomb repulsion. The density wave state has a gapped energy spectrum at $n=2$ and yields a single doubly-degenerate pocket upon doping to $n>2$. The intertwinement of density wave and superconductivity and the quasiparticle spectrum in the density wave state are consistent with experimental observations.
\end{abstract}

\maketitle

\section{Introduction}

Recently superconductivity was discovered near a correlated insulator state in bilayer graphene with a small twist angle $\theta \approx 1.1^\circ$ \cite{exp1,exp2}, where the moir\'e pattern creates a superlattice with a periodicity of about 13\,nm. A correlated insulating state is found at the filling of $n=2$ electrons per supercell ($n=0$ is the charge neutrality point). Electron or hole doping away from $n=2$ by electrostatic gating leads to a superconducting dome, similar to cuprates. Insulating states are also found in trilayer graphene with moir\'e superlattice \cite{Wang}.  The nature of superconducting and insulating states in graphene superlattices are now under intensive theoretical study \cite{{Yuan},{Xu},{Volovik},{Po},{Zhang}, {Kivelson},{Philips},{Baskaran},{Lee},{Yang}}.

For $\theta \approx 1.1^\circ$, the low-energy mini-band of twisted bilayer graphene has a narrow bandwidth of 10\,meV scale \cite{Nam,TB1,TB2,TB3,TB4,TB5,DFT1,DFT2,Nori2,Neto,Neto2,MacDonald, Mele}. However, this energy scale is still much larger than the energy gaps of the superconducting and insulating states, which are on the order of 1\,K. Moreover, resistivity shows metallic behavior above 4\,K. This is rather different from the case of a Mott insulator in the strong coupling limit, which would become insulating at much higher temperature. Based on these considerations, in this work, we take a weak coupling approach to study ordered states driven by electron correlation in twisted bilayer graphene.

\begin{figure}
\centering
\includegraphics[width=0.9\hsize]{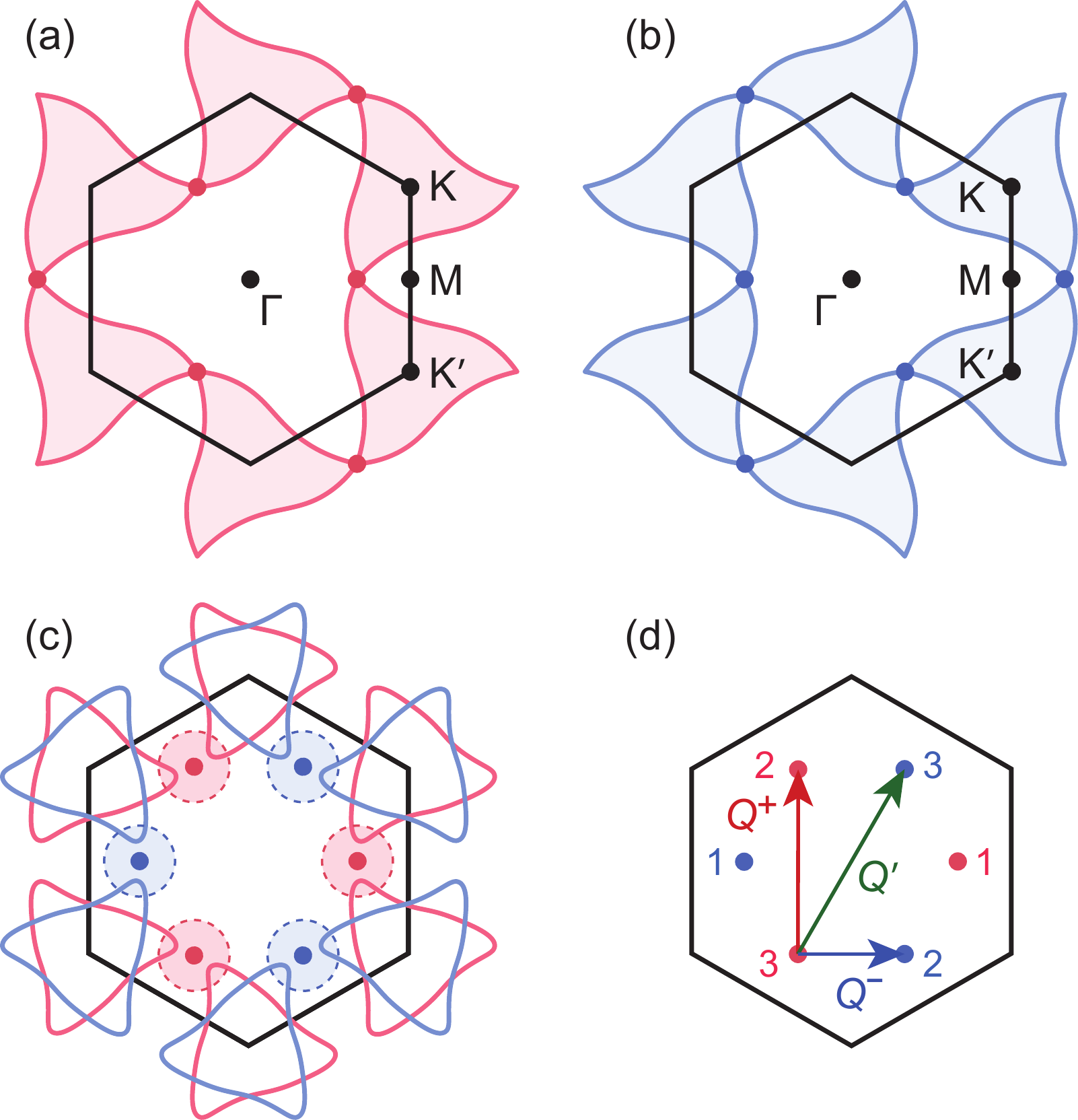}
\caption{
(a), (b) Two Fermi surfaces at the Van Hove energy from different valleys shown in red and blue, reproduced from Moon and Koshino's band structure calculation for twisted bilayer graphene with $\theta=2^\circ$~\cite{Kim}.  Shaded areas are filled %around the filling of $n=2$
and VHS appear at the points, where two Fermi surfaces encircling $K$ and $K'$ touch.  Each Fermi surface has $C_3$ symmetry about $\Gamma$, $K$, and $K'$. The  Fermi surfaces in (a) and (b) are related by $C_2$ rotation with respect to an in-plane axis along $\Gamma$--$K$.
(c) Two Fermi surfaces slightly away from the Van Hove energy.  DOS is larger near the VHS points (hot spots), where electron interaction predominates.  We assign patches (circles with dashed lines) centered at hot spots.
(d) Three inequivalent wave vectors ($Q^+$, $Q^-$, and $Q'$), along with symmetry-related ones (not shown), connect various pairs of hot spots.
}
\label{fig:Fermi_surface}
\end{figure}

While details of the band structure remain to be fully sorted out, a number of prominent features of the normal state fermiology are robust and noteworthy. First, at small twist angle, the two valleys of graphene have negligibly small single-particle hybridization and give rise to two separate Fermi surfaces that intersect each other in the mini Brillouin zone \cite{MacDonald, PRL2016, Kim}.
Second, as the carrier density increases, Fermi pockets associated with a given valley first appear around the Dirac points at  $K$ and $K'$ in the mini Brillouin zone, then these $K$ and $K'$ pockets merge at a saddle point on the $\Gamma$--$M$ line to become a single pocket centered at $\Gamma$. The saddle point associated with this Lifshitz transition has a Van Hove singularity (VHS) with a logarithmic divergence of the density of states (DOS). Third, realistic band structure calculations \cite{Kim} show that near the Van Hove energy the Fermi surfaces of different valleys contain nearly parallel segments and hence are nearly nested, see Fig.~\ref{fig:Fermi_surface}. Such Fermi surface nesting strongly enhances density wave fluctuations.

When the Fermi energy crosses the Van Hove energy, a conversion between electron and hole carriers is expected and indeed observed from the sign change of Hall resistance as a function of doping for $\theta=2^\circ$ \cite{Kim} and $\theta=1.8^\circ$ \cite{PRL2016}. Remarkably, this sign change occurs at the filling $n=2$, indicating that the Fermi energy is very close to VHS. VHS in twisted graphene layers was also detected from pronounced peaks in DOS in STM measurements \cite{Andrei}. As the twist angle becomes smaller, the energy separation between VHS in conduction and valence bands is found to decrease rapidly from 430\,meV at $\theta = 3.4^\circ$ to 82\,meV at $1.79^\circ$ and 12\,meV at $1.16^\circ$.  The strong reduction of bandwidth is expected to magnify the effects of electron correlation. This understanding is consistent with the fact that correlated insulator and superconducting states are found for $\theta = 1.1^\circ$, but not for $\theta =1.8^\circ$ and $2^\circ$. In the following, we take the Fermi surface with good nesting condition and in proximity to VHS as a starting point and study its instabilities in the presence of electron interactions.

Due to the divergent DOS at VHS, electron interaction predominates in patches of the Brillouin zone around saddle points, or ``hot spots''.
When multiple hot spots are present at a given energy, various scattering processes among them may interfere with each other, leading to intertwined density wave and superconducting instabilities.
Such hot-spot models were studied with renormalization group (RG) approach in the context of cuprates \cite{Schulz,Dzyaloshinskii,Lederer,Furukawa,LeHur}, and recently, by Nandkishore, Levitov, and Chubukov in the context of doped monolayer graphene \cite{Nandkishore}.

In this paper, we study interaction-driven ordering instabilities of twisted bilayer graphene around the filling $n=2$ using RG by patching the Brillouin zone where the DOS is considerably larger than other parts.
Our RG analysis shows how the electron interaction changes as the energy scale is reduced. Nontrivial RG flows of the coupling constants are found as a consequence of the nesting of Fermi surfaces.
Susceptibility calculations reveal the possibility of various superconducting and spin/charge density-wave states at low temperature. When Coulomb repulsion is the dominant interaction, $d$/$p$-wave superconductivity and charge/spin density wave at a particular nesting wave vector emerge as two leading instabilities. The density-wave state is found to have a gapped energy spectrum at $n=2$ and yields a single doubly-degenerate pocket upon doping to $n>2$.

\section{Model}

We set up a model to analyze the electron interaction effect in twisted bilayer graphene around the filling $n=2$.  Our analysis of the interaction-driven instabilities focuses on hot spots, which dominates in the DOS.   The hot spots are patches on the Brillouin zone.  The Fermi surface nesting in the hot spots may potentially lead to ordering instabilities.  In the following, we introduce a hot-spot model for twisted bilayer graphene and then the notion of Fermi surface nesting in hot spots.

\subsection{Hot-spot model}
\label{sec:hot-spot}

To consider the electron interaction effect and resultant ordering instabilities near the filling $n=2$, we focus on hot spots in the Brillouin zone which possess significantly larger electronic spectral weights compared to the other region.  Such hot spots are obtained by patching the Brillouin zone around the saddle points.
The patches are labeled by $\tau=\pm1$ for Fermi surfaces from the two valleys, $\sigma$ for spins, and $\alpha=1,\ldots,3$ for the patches of a given valley (Fig.~\ref{fig:Fermi_surface}). We denote the position of a patch center by $\bm{k}_{\alpha\tau}$.  The three inequivalent wave vectors connecting the patch centers are defined by $\bm{Q}^{+}=\bm{k}_{\alpha\tau}-\bm{k}_{\beta\tau}$ (intravalley), $\bm{Q}^{-}=\bm{k}_{\alpha\tau}-\bm{k}_{\beta\tau'}$ and $\bm{Q}'=\bm{k}_{\alpha\tau}-\bm{k}_{\alpha\tau'}$ (intervalley) ($\tau\neq\tau'$, $\alpha\neq\beta$), see Fig.~\ref{fig:Fermi_surface}(d).

\begin{figure}
\centering
\includegraphics[width=.9\hsize]{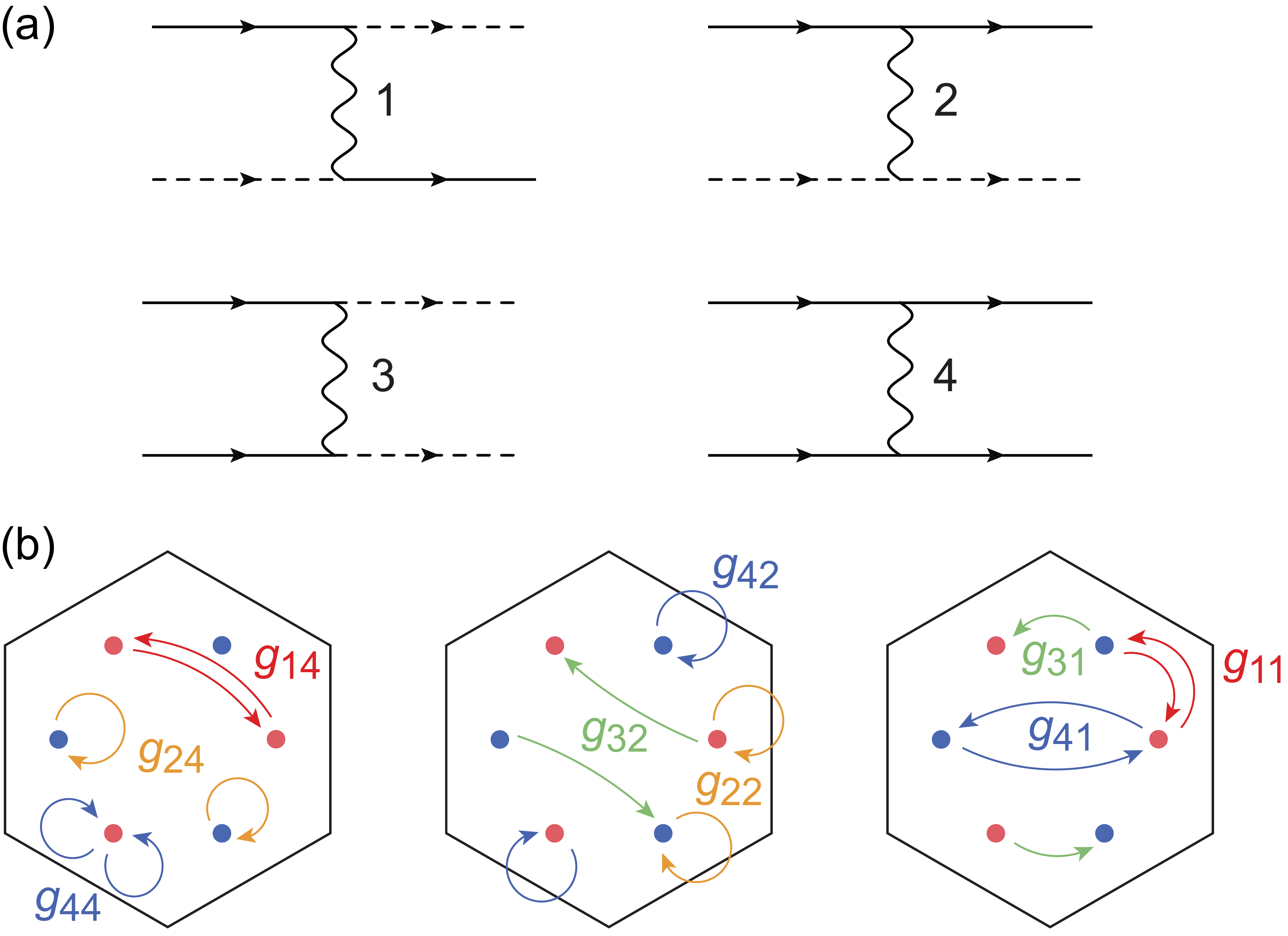}
\caption{
(a) Diagrammatic representation of scattering processes. The four diagrams describe the change of either saddle point or valley index, where solid and dashed lines correspond to electron propagators with different indices.  %Since a scattering affects both saddle point and valley degrees of freedom, there are in total sixteen different scattering processes.
(b) Nine momentum-conserving scattering processes out of sixteen distinct scattering processes $g_{ij}$. Hexagons are the Brillouin zone boundaries and the dots are located at saddle points with two colors corresponding to the two valleys.
There are three types in the interactions (from left to right): intravalley, intervalley density, and intervalley exchange interactions.
}
\label{fig:scattering}
\end{figure}

Electron interaction is treated as scattering among the patches.
By analogy with the $g$-ology model in one-dimensional physics \cite{1d_1,1d_1-1,1d_2} and Shankar's RG approach \cite{Shankar}, we write down the general interaction Hamiltonian compatible with lattice rotational symmetry
\begin{align}
\label{eq:interation}
H_\text{int} = \frac{1}{2} \sum_{i,j=1}^{4} \sum_{\substack{\alpha_1,\ldots,\alpha_4 \\ \tau_1,\ldots,\tau_4}} \sum_{\sigma\sigma'}
g_{ij}\psi^\dagger_{\alpha_1\tau_1\sigma} \psi^\dagger_{\alpha_2\tau_2\sigma'} \psi_{\alpha_3\tau_3\sigma'} \psi_{\alpha_4\tau_4\sigma}.
\end{align}
Here the patch indices satisfy $\alpha_1=\alpha_3\neq\alpha_2=\alpha_4$ $(i=1)$, $\alpha_1=\alpha_4\neq\alpha_2=\alpha_3$ $(i=2)$, $\alpha_1=\alpha_2\neq\alpha_3=\alpha_4$ $(i=3)$, and $\alpha_1=\alpha_2=\alpha_3=\alpha_4$ $(i=4)$.  The valley indices $\tau_1,\ldots,\tau_4$ obey the same rule, associated with $j$.
This rule is diagrammatically shown in Fig.~\ref{fig:scattering}(a). The interaction describes sixteen independent scattering processes with coupling constants $g_{ij}$
In the following analysis, we consider the momentum-conserving processes depicted in Fig.~\ref{fig:scattering}(b). Scattering processes related to these ones by lattice symmetry are not shown.
Note that $g_{i3}$, $g_{12}$, $g_{21}$, and $g_{34}$ do not generally conserve crystal momentum since the patches are located away from the Brillouin zone boundary (see Fig.~\ref{fig:scattering_2}). Umklapp processes are allowed only when the hot spots are located at special momenta. The analysis for that case includes more or all $g_{ij}$, and is presented in Appendix~\ref{sec:generalized_model}.  Among the nine momentum-conserving terms, $g_{11}$, $g_{14}$, $g_{22}$, $g_{24}$, $g_{44}$ are associated with forward scattering processes, and $g_{31}$, $g_{32}$, $g_{41}$, $g_{42}$ are BCS scattering processes involving two electrons with opposite momenta.

The nine momentum-conserving terms $g_{ij}$ in the interaction Hamiltonian Eq.~(\ref{eq:interation}) can be divided into three groups with different index $j$. For $g_{i4}$ terms, all four operators belong to one valley $(\tau_1=\tau_2=\tau_3=\tau_4)$, thus describing intravalley interactions.  $g_{i2}$ terms are the product of two spin- and valley-conserving fermion bilinear operators that are associated with two different valleys; we call them intervalley density interactions. $g_{i1}$ terms are the product of two spin-conserving but valley-flipping fermion bilinear operators; we call them intervalley exchange interactions.

We now discuss the microscopic origin of these interaction terms and how the coupling constants $g_{ij}$ should in principle be determined. First, when projected to the lowest mini-band, the long-range part of Coulomb interaction generates intra- and intervalley density interaction $g_{i2}$, $g_{i4}$, while the short-range part of Coulomb interaction on graphene's lattice scale generates intervalley exchange interaction $g_{i1}$ involving large momentum transfer. Since Wannier functions in twisted bilayer graphene extend over tens of nanometers, the long-range part of Coulomb interaction is expected to dominate. Based on this factor alone, one would expect density interactions $g_{i2}$, $g_{i4}$ to be orders-of-magnitude larger than the intervalley exchange interaction $g_{i1}$. On the other hand, it should be noted that the long-range Coulomb interaction strength is reduced by screening from excited bands that span a wide range of energies from $\sim 10\,\mathrm{meV}$ up to the bandwidth of graphene layers $\sim 10\,\mathrm{eV}$. The process of integrating out these excited bands as well as those states of the lowest band outside the patches will significantly renormalize the coupling constants to be used in our patch theory. Moreover, their values are also affected by electron-phonon coupling.  Since typical phonon energy in graphene is much larger than the mini-band width, it is reasonable to integrate out the phonons to obtain phonon-mediated electron-electron attraction \cite{Kivelson}, which renormalizes the values of coupling constants.

In this work, we treat the bare values of $g_{ij}$ as phenomenological parameters and calculate flows of these coupling constants under RG to the one-loop order.  Strictly speaking, such a perturbative RG analysis is only legitimate for weak coupling. However, instabilities toward superconductivity and/or density waves are found within the weak-coupling regime (see below), which justifies the one-loop RG analysis.

\subsection{Susceptibilities and Fermi surface nesting}

Fermion loops in the RG calculation are associated with bare susceptibilities in the particle-hole and particle-particle channels:
\begin{gather}
\chi^\text{ph} (\bm{q}_-,\omega)
= \int_{\bm{k}\in\text{patch}} \frac{f(\epsilon^\tau_{\bm{k}+\bm{k}_{\alpha\tau}})-f(\epsilon^{\tau'}_{\bm{k}+\bm{k}_{\beta\tau'}})}{\omega-\epsilon^\tau_{\bm{k}+\bm{k}_{\alpha\tau}}+\epsilon^{\tau'}_{\bm{k}+\bm{k}_{\beta\tau'}}}, \\
\chi^\text{pp} (\bm{q}_+,\omega)
= \int_{\bm{k}\in\text{patch}} \frac{f(\epsilon^\tau_{\bm{k}+\bm{k}_{\alpha\tau}})-f(-\epsilon^{\tau'}_{-\bm{k}+\bm{k}_{\beta\tau'}})}{\omega-\epsilon^\tau_{\bm{k}+\bm{k}_{\alpha\tau}}-\epsilon^{\tau'}_{-\bm{k}+\bm{k}_{\beta\tau'}}},
\end{gather}
where $\bm{q}_\pm=\bm{k}_{\alpha\tau}\pm\bm{k}_{\beta\tau'}$, $f(\epsilon)$ is the Fermi distribution, and $\epsilon^\tau_{\bm{k}}$ is the energy dispersion of valley $\tau$ with $\epsilon=0$ on the Fermi surface.

There are in total eight susceptibilities in the particle-particle and particle-hole channels at various wave vectors:
\begin{gather}
\label{eq:chi_i}
\chi_{0+}=\chi^\text{pp}(Q',\omega), \quad \chi_{0-}=\chi^\text{pp}(0,\omega), \\
\chi_{1s}=\chi^\text{ph}(Q^s,\omega), \\
\chi_{2+}=\chi^\text{ph}(0,\omega), \quad \chi_{2-}=\chi^\text{ph}(Q',\omega), \\
\chi_{3s}=\chi^\text{pp}(Q^{-s},\omega),
\label{eq:chi_f}
\end{gather}
where $s=+,-$ correspond to intra- and intervalley components, respectively.

Among these susceptibilities, $\chi_{2+}(\omega=0)$ is the DOS within a patch at Fermi energy $\rho_0$.  $\chi_{0-}$ is the $Q=0$ Cooper-pair susceptibility, which involves patches at opposite momenta, belonging to the different valleys. Regardless of Fermi surface geometry, $\chi_{0-}$ exhibits a logarithmic divergence
$\chi_{0-}(\omega) = \frac{\rho_0}{4} \ln\frac{\Lambda}{\omega}$ in the presence of time-reversal symmetry, where $\Lambda$ is the high-energy cutoff and depends on the patch size.

Besides the BCS channel, susceptibilities in other channels may find divergences when Fermi surfaces are perfectly nested: $\epsilon^\tau_{\bm{k}+\bm{k}_{\alpha\tau}}=-\epsilon^{\tau'}_{\bm{k}+\bm{k}_{\beta\tau'}}$ for the particle-hole channels and $\epsilon^\tau_{\bm{k}+\bm{k}_{\alpha\tau}}=\epsilon^{\tau'}_{-\bm{k}+\bm{k}_{\beta\tau'}}$ for the particle-particle channels. In general, the interplay between BCS and nesting-related interactions can lead to nontrivial RG flows of coupling constants \cite{Shankar,Chubukov}.

As shown in Fig.~\ref{fig:Fermi_surface}, the Fermi surfaces associated with different valleys have nearly parallel segments connected by the following nesting vectors: $Q^-$ and $Q'$ connect occupied states of one valley and unoccupied states of another, while $Q^+$ connects occupied states around $K$ and those around $K'$ associated with the same valley; see Fig.~\ref{fig:Fermi_surface}(d). Such Fermi surface nesting strongly enhances three types of bare susceptibilities: intervalley charge or spin density wave susceptibility at $Q^-$ and $Q'$ (particle-hole channel), and intervalley pair density wave at $Q^+$ (particle-particle channel).

In the ideal case where the Fermi surface is perfectly nested (see Appendix~\ref{sec:nesting} for the detailed discussion), these susceptibilities are logarithmically divergent like the BCS channel.
When the Fermi surface is nearly nested, the logarithmic frequency energy dependence still holds approximately within a range $\Lambda_0 < \omega \leq \Lambda$, but the divergence in the $\omega \rightarrow 0$ limit is cutoff below a smaller energy scale $\Lambda_0$ associated with the deviation from perfect nesting.

Finally, when the VHS lies close to the Fermi surface, scatterings among states near VHS points receive particularly large RG corrections because of the large spectral weight. This justifies our use of patch RG approach.  When the VHS lies exactly on the Fermi surface, the DOS at $\omega=0$ is logarithmically divergent and thus leads to an additional log divergence in susceptibilities, see discussion in Appendix~\ref{sec:chi_VHS}.

\section{RG analysis}
\label{sec:RG}

\begin{figure}
\centering
\includegraphics[width=\hsize]{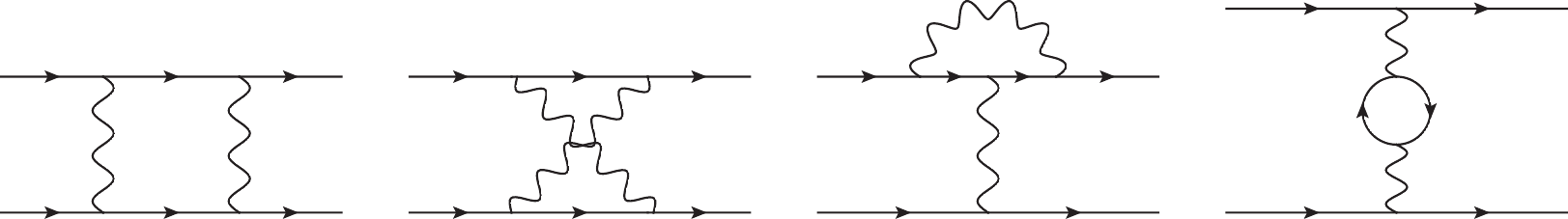}
\caption{
One-loop corrections to the coupling constants. The solid lines represents the fermion propagators and the wavy lines correspond to interactions. The leftmost diagram involves a particle-particle loop, and the other three diagrams have particle-particle loops.
}
\label{fig:diagram_loop}
\end{figure}

Loop corrections to the coupling constants suffer from divergences due to Fermi surface nesting and divergent DOS at the Van Hove energy, which are to be cured with the RG method.  We consider corrections to one-loop order (Fig.~\ref{fig:diagram_loop}).
In the hot-spot model, each loop corrections is associated with the susceptibilities Eqs.~\eqref{eq:chi_i}--\eqref{eq:chi_f}.
Since the Cooper-pair susceptibility $\chi_{0-}$ always gives the leading divergence regardless of the Fermi surface geometry, we set the RG scale by $y\equiv \chi_{0-}(\epsilon)$.
In the Wilsonian RG procedure, we integrate out high-energy modes and rescale the remaining low-energy modes as we increase the RG scale $y$, which is qualitatively similar to lowering the temperature $T$ down from $\Lambda$.
The other susceptibilities are measured with respect to $y$, parameterized by
\begin{gather}
d_{as}(y)=\frac{d\chi_{as}}{dy}.  \label{d}
\end{gather}
By definition, $d_{0-}=1$ always holds.

\subsection{RG equations}

The RG equations for the coupling constants involve the parameters $d_{as}(y)$ defined in Eq.~\eqref{d} as a function of the RG scale $y$.  In general, $d_{as}(y)$ depends on $y$, except when the corresponding susceptibility  $\chi_{as}(y)$ diverges similarly to the BCS susceptibility. This occurs in the presence of Fermi surface nesting. Then the corresponding density-wave channel has a divergent susceptibility, and hence $d_{as}(y) = d_{as}$ is a constant less than or equal to $1$ ~\cite{Furukawa,LeHur,Nandkishore}.

In an ideal case for nesting where Fermi surface comprises of corner-sharing triangles (see Fig.~\ref{fig:FS_simplified}), perfect nesting occurs simultaneously in three channels: three susceptibilities of the intervalley type,  $\chi_{1-}(Q^-)$, $\chi_{2-}(Q')$, and $\chi_{3-}(Q^+)$, are all logarithmically divergent similar to $\chi_{0-}$, so that $d_{1-}, d_{2-}$ and $d_{3-}$ are nonzero constants. In contrast, none of the intravalley susceptibilities $\chi_{a+}$ is divergent. Thus $d_{a+}(y)$ decay as $y^{-1}$ (away from the Van Hove energy) or $y^{-1/2}$ (at the Van Hove energy; see Appendix~\ref{sec:chi_VHS}), and hence become negligible at large $y$. We neglect the subleading terms $d_{a+}$ in the following analysis. (RG equations for a generalized model with $d_{a+}$ and additional interaction terms are presented in Appendix~\ref{sec:generalized_model}.)

As shown in Fig.~\ref{fig:Fermi_surface}, the Fermi surface of twisted bilayer graphene is nearly (but not perfectly) nested. In this case, the intervalley susceptibilities $\chi_{a-}$ ($a=1,2,3$) still have a logarithmic dependence on energy from $\Lambda$ down to a smaller energy $\Lambda_0$.
Equivalently, the parameter $d_{a-}(y)$ is approximately constant within the corresponding range of the RG scale $0\leq y<y_0$. In the following, we shall analyze the RG flow within this energy range of interest, where the Fermi surface is regarded as nested.

For $d_{a+}=0$, we obtain the RG equations for the nine momentum-conserving coupling constants as follows:
\begin{align}
\label{eq:RG_i}
& \dot{g}_{14} = \dot{g}_{24} = \dot{g}_{44} =0, \\
&\begin{aligned}
\label{eq:RG22}
\dot{g}_{22}
=& -d_{3-}(g_{11}^2+g_{22}^2) +d_{1-}(g_{22}^2+g_{32}^2),
\end{aligned}\\
&\begin{aligned}
\label{eq:RG32}
\dot{g}_{32}
=& -(g_{31}^2+g_{32}^2+2g_{31}g_{41}+2g_{32}g_{42}) \\
 & +2d_{1-}g_{22}g_{32},
\end{aligned}\\
&
\label{eq:RG42}
\dot{g}_{42}
= -(2g_{31}^2+2g_{32}^2+g_{41}^2+g_{42}^2) +d_{2-}g_{42}^2, \\
&\begin{aligned}
\dot{g}_{11}
=& -2d_{3-}g_{11}g_{22} \\
 & +2d_{1-}(g_{11}g_{22}-g_{11}^2+g_{31}g_{32}-g_{31}^2),
\end{aligned}\\
&\begin{aligned}
\dot{g}_{31}
=& -2(g_{31}g_{32}+g_{31}g_{42}+g_{32}g_{41}) \\
 & +2d_{1-}(g_{11}g_{32}+g_{22}g_{31}-2g_{11}g_{31}),
\end{aligned}\\
&
\dot{g}_{41}
= -2(2g_{31}g_{32}+g_{41}g_{42}) +2d_{2-}(g_{41}g_{42}-g_{41}^2).
\label{eq:rg_f}
\end{align}
We use the convention $\dot{g}\equiv dg/dy$.

Equations~\eqref{eq:RG_i}--\eqref{eq:rg_f} show how different coupling constants change under RG. Among them, the intervalley interactions $g_{22}$ and $g_{11}$ involve two patches not related by time-reversal symmetry, hence receive corrections solely from scattering processes related to intervalley nesting. $g_{32}$, $g_{42}$, $g_{31}$, and $g_{41}$ involve two patches related by time-reversal symmetry, hence receive corrections from both BCS and nesting-related processes. The intravalley interactions $g_{i4}$ do not flow because they do not participate in either process.

Details of the RG flow in the nine-dimensional parameter space are complicated and can in general be acquired numerically. (See Appendix~\ref{sec:RG_analytic} for the simplest case without nesting, where the analytic solution is obtained.) Nonetheless, its general feature can be understood easily:  BCS corrections decrease repulsive interactions under RG, while nesting-related corrections tend to increase repulsive interactions in the corresponding channels.  This important trend is a useful guideline to understand the behavior of the RG flow, which we present later.

\subsection{Ordering instabilities}

To analyze various possible instabilities, we consider susceptibilities associated with $s$- and $d$-wave spin-singlet superconductivity  ($s$-SC etc.), $p$- and $f$-wave spin-triplet superconductivity, CDW, SDW, and pair density wave (PDW).
Both $p$- and $d$-wave pairings have two degenerate components: $(p_x, p_y)$ and $(d_{xy}, d_{x^2-y^2})$; see Appendix~\ref{sec:general_analysis} for detail.
Three different wave vectors, $Q^+$, $Q^-$, and $Q'$ associated with density-wave orders are indicated by superscripts $+$, $-$, and $'$, e.g., SDW$^-$ and CDW$'$.

When only the intervalley Fermi surface nesting is considered, the relevant instabilities are superconductivity, CDW/SDW at wave vectors $Q^-$ and $Q'$, and PDW at $Q^+$.  An occurrence of an instability is indicated by a divergence of the corresponding susceptibility. The susceptibility  are obtained by an RPA-like resummation \cite{Chubukov}
\begin{align}
\chi_\eta(y) &= \chi_\eta^0(y) - \chi_\eta^0(y) V_\eta(y) \chi_\eta^0(y) + \ldots \nonumber\\
&= \frac{\chi_\eta^0(y)}{1+V_\eta(y)\chi_\eta^0(y)},
\end{align}
where $\eta$ is used to label various ordering susceptibilities. $\chi_\eta^0(y)$ is the bare susceptibility in the absence of interaction and $V_\eta(y)$ is the effective interaction strength associated with the ordering.

By a straight-forward diagrammatic calculation, we find $V_\eta$ for various ordering channels as follows: %\cite{SM}
\begin{gather}
\label{eq:interaction_strength_1}
% \begin{gathered}
V_{s,d\text{-SC}} = 2(g_{42}+g_{41} \pm g_{32} \pm g_{31}) \quad\text{(singlet SC)}, \\
\label{eq:interaction_strength_2}
V_{p,f\text{-SC}} = 2(g_{42}-g_{41} \mp g_{32} \pm g_{31})　\quad\text{(triplet SC)}, \\
% V_{\text{CDW}^+} = 4(g_{14}+g_{32})-2(g_{24}+g_{31}), \\
\label{eq:interaction_strength_CDW-}
V_{\text{CDW}^-} = 4 (g_{11}+g_{31})-2(g_{22}+g_{32}), \\
\label{eq:interaction_strength_CDW'}
V_{\text{CDW}'} = 4g_{41}-2g_{42}, \\
% V_{\text{SDW}^{+}} = -2(g_{24}+g_{31}), \\
\label{eq:interaction_strength_SDW-}
V_{\text{SDW}^{-}} = -2(g_{22}+g_{32}), \\
\label{eq:interaction_strength_SDW'}
V_{\text{SDW}'} = -2g_{42}, \\
V_{\text{PDW}^+} = 2(-g_{11}+g_{22}).
\label{eq:interaction_strength_f}
% V_{\text{PDW}^-} = 2(-g_{14}+g_{24}), \\
% V_{\text{PDW}'} = 2g_{44}.
% \end{gathered}
\end{gather}
Detailed description and derivation of interaction strengths and susceptibilities are found in Appendix~\ref{sec:general_analysis}.

When the parameters $d_{as}$ are constant, to the leading order in $y$, $\chi_\eta(y)$ is written as
\begin{equation}
\label{eq:susceptibility_y}
\chi_\eta(y) = \frac{d_{as} y}{1+V_\eta(y)d_{as} y}.
\end{equation}
The susceptibility diverges at $1+V_\eta(y_c)\chi_\eta(y_c)=0$, leading to an instability at $y_c$.  An instability occurs only if the interaction strength is attractive: $V_\eta<0$.   In mean-field theory, the interaction strength $V_\eta(y)$ is treated as a constant determined by the bare values of the coupling constants $g_{ij}$. Then the instability temperature is given by
%\begin{equation}
$T_{\eta} = \Lambda \exp[-(c_p d_{as} |V_\eta|)^{-1/p}]$,
%\end{equation}
for $y=c_p \ln^p(\Lambda/\epsilon)$ ($c_p>0$) with $p=1$ away from the VHS or $p=2$ at the VHS. Our analysis here further takes into account the RG-scale dependence of coupling constants $g_{ij}$ and hence the interaction strength $V_\eta(y)$. We shall see that the running coupling constants leads to results beyond mean-field theory.

\subsection{Intertwined superconductivity and density waves}
\label{sec:intertwine_main}

\begin{figure*}
\centering
\includegraphics[width=0.8\hsize]{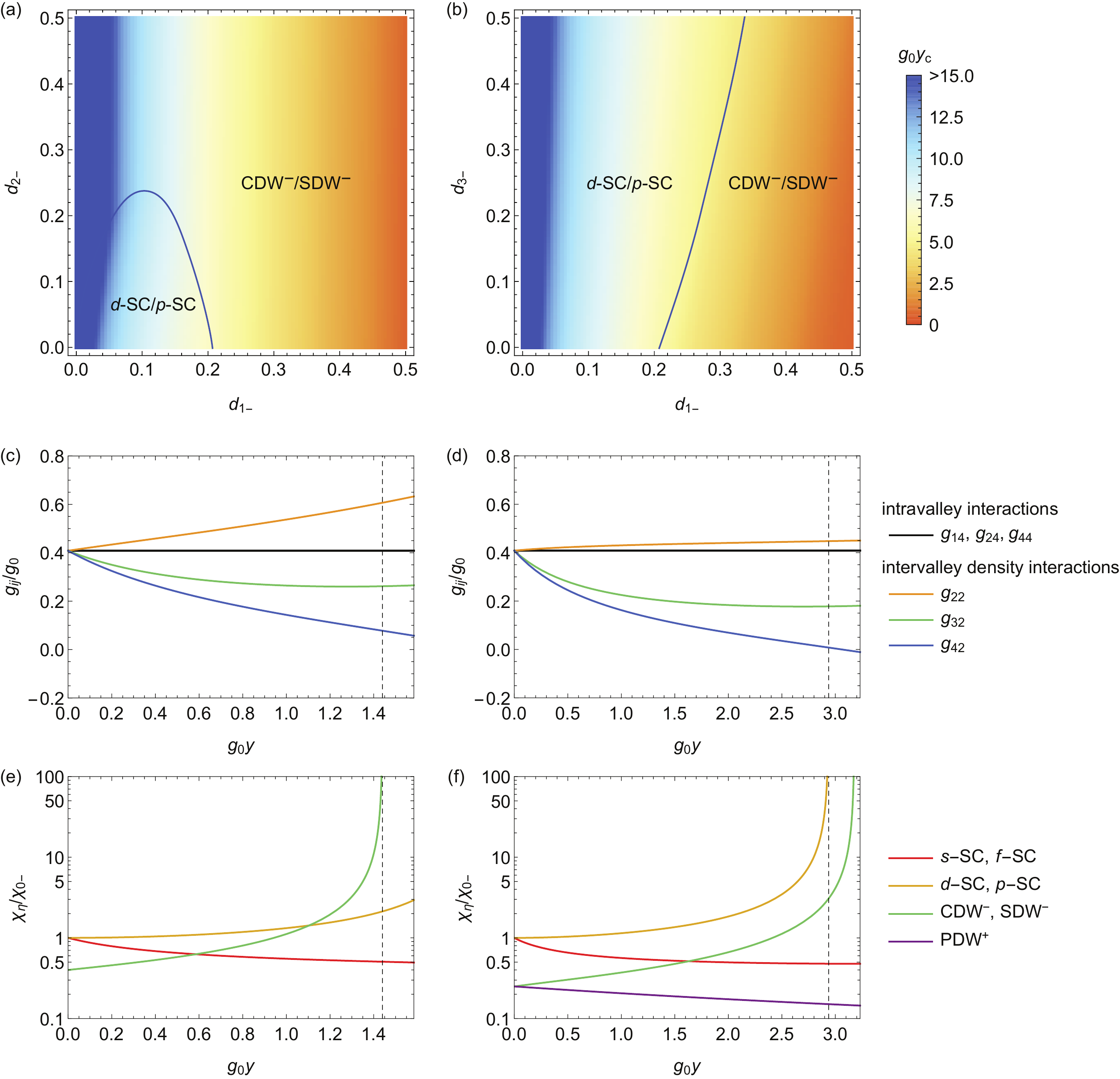}
\caption{
Phase diagrams with the density-density interactions.
(a), (b) Phase diagrams obtained by the RG analysis in the presence of the density-density interactions ($g_{14}=g_{24}=g_{44}=g_{22}=g_{32}=g_{42}$).  Two parameters for nesting are varied in each phase diagram: (a) $d_{1-}$ and $d_{2-}$ ($d_{3-}=0$); and (b) $d_{1-}$ and $d_{3-}$ ($d_{2-}=0$).  Colors represent critical RG scales for instabilities: warm (cool) colors correspond to high (low) energy and temperature.  The solid lines represent phase boundaries, obtained in the range of $g_0y_c\leq15$.  There are CDW$'$/SDW$'$ and normal regions in the vicinity of $d_{1-}=0$ (not shown), but critical temperatures are extremely low for those instabilities ($g_0y_c>15$).
(c), (d) Running coupling constants and (e), (f) susceptibilities for possible orders.  We show the results for two cases with different strength of nesting: (c), (e) $d_{1-}=0.4$ and (d), (f) $d_{1-}=d_{3-}=0.25$.  The vertical dashed lines indicate the positions where instabilities occur.
}
\label{fig:phase_dd}
\end{figure*}

Since the intervalley exchange interaction is likely smaller than the density-density interaction, it is instructive to first analyze cases only with the density-density interactions.  For simplicity, we set the strengths of all density-density interactions equal $(g_{14}=g_{24}=g_{44}=g_{22}=g_{32}=g_{42}>0)$.
In the absence of the exchange interactions, some susceptibilities become degenerate as we see from Eqs.~\eqref{eq:interaction_strength_1}--\eqref{eq:interaction_strength_f}: $s$-SC and $f$-SC, $d$-SC and $p$-SC, and CDW and SDW at each wave vector. Such degeneracy results from the fact that each valley has its own charge conservation and spin rotation symmetry. Similar degeneracy also occurs in exciton insulators when only long-range Coulomb interaction is considered \cite{Halperin}.

With this choice of coupling constants, a mean-field analysis does not find any superconducting instability because the pairing interactions shown in Eqs.~\eqref{eq:interaction_strength_1} and \eqref{eq:interaction_strength_2} are zero.
In the presence of Fermi surface nesting, a density wave instability is found in mean-field analysis, whose wave vector is $Q^-$ if $d_{1-}(g_{22}+g_{32})>d_{2-}g_{42}$, and $Q'$ vice versa.

Our RG analysis including the scale dependence of the coupling constants $g_{ij}$ finds qualitatively different results.
Figures~\ref{fig:phase_dd}(a) and (b) are the phase diagrams from the one-loop RG analysis on the $(d_{1-}, d_{2-})$ plane with $d_{3-}=0$ and the $(d_{1-}, d_{3-})$ plane with $d_{2-}=0$, respectively.  First, in both cases, $d$- or $p$-wave superconductivity is found for small $d_{1-}$, which is absent in mean-field theory.
Second, the $Q^-$ density-wave state is far more dominant than the $Q'$ density-wave state: it already occurs at very small nesting parameter $d_{1-}$, even when the Fermi surface nesting condition is much stronger at wave vector $Q'$.

To understand why the $Q^-$ density wave and $d$/$p$-wave superconductivity emerge as leading instabilities, we examine the flow of intervalley density interactions $g_{22}$, $g_{32}$, and $g_{42}$.  Since $g_{32}$ and $g_{42}$ are associated with BCS scattering processes, they receive renormalization even without nesting and decrease under RG when their initial values are repulsive, as shown in Figs.~\ref{fig:phase_dd}(c), (d). In contrast,  since $g_{22}$ is associated with a forward scattering process, it is marginal without nesting.  According to the RG equation Eq.~\eqref{eq:RG22}, Fermi surface nesting in the particle-hole channel ($d_{1-} >0$) increases $g_{22}$ under RG.
Therefore, in the presence of Fermi surface nesting, only $g_{22}$ grows without suppression from the BCS process and thus strongly enhances the $Q^-$ density wave fluctuation, making it dominate over the $Q'$ density wave.

Although $g_{32}$ and $g_{42}$ both decrease under RG, the former decreases slower because BCS process and density wave nesting at wave vector $Q^-$ tend to renormalize $g_{32}$ in the opposite way; see Eq.~\eqref{eq:RG32}.  Therefore, a negative $g_{42}-g_{32}<0$ is generated and its magnitude grows under RG. As shown in Eqs.~\eqref{eq:interaction_strength_1} and \eqref{eq:interaction_strength_2}, this attraction provides pairing interaction for both $d$-SC and $p$-SC and thus enhances these superconducting susceptibilities, see Fig.~\ref{fig:phase_dd}(f).
The attractive pairing interaction should be stronger than that for the $Q^-$ density wave.   Fermi surface nesting in the particle-particle channel ($d_{3-}>0$) assists superconductivity in that it suppresses the increase of $g_{22}$.  Finite nesting in the particle-particle channel yields nonzero PDW susceptibility, but the interaction is repulsive for the PDW$^+$ fluctuation; see Eq.~\eqref{eq:interaction_strength_f}.

We conclude that in the presence of repulsive intervalley density interactions, the two leading instabilities are charge/spin density wave at wave vector $Q^-$ and $p$/$d$-wave superconductivity. When the Fermi surface nesting in the particle-hole channel is strong (weak), the density wave state (superconductivity) is favored.

\begin{figure*}
\centering
\includegraphics[width=\hsize]{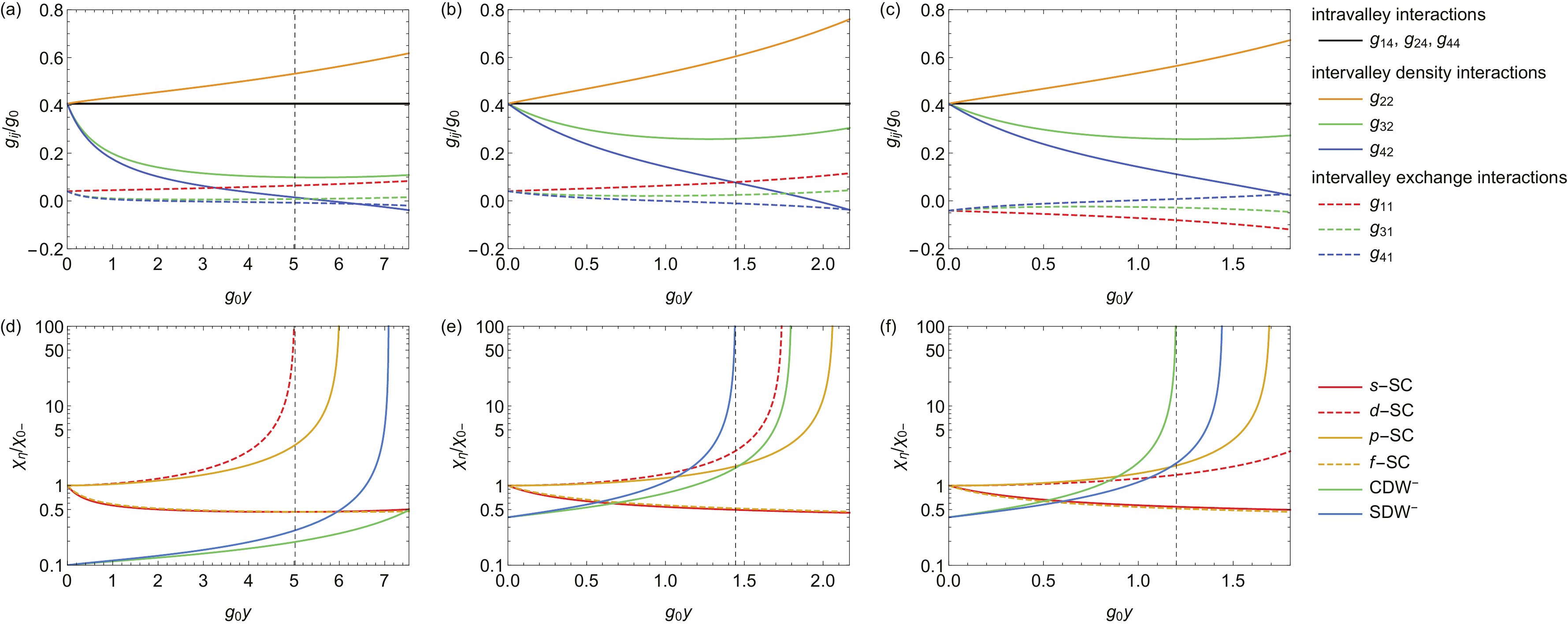}
\caption{
Effect of the exchange interactions.
(a)--(c) Running coupling constants and (d)--(f) susceptibilities for possible orderings.  We choose the strengths of density-density interactions $g_{14}=g_{24}=g_{44}=1$ (intravalley) and $g_{22}=g_{32}=g_{42}=1$ (intervalley).
Fermi surface nesting is weak in (a) and (d) with $d_{1-}=0.1$; and strong in (b), (c), (e), and (f) with $d_{1-}=0.4$.
Finite exchange interactions lift the degeneracies of susceptibilities; cf.\ Figs.~\ref{fig:phase_dd}(e), (f).
We choose the exchange interactions to be initially repulsive ($g_{11}=g_{31}=g_{41}=0.1$) in (a), (b), (d), and (e); and attractive ($g_{11}=g_{31}=g_{41}=-0.1$) in (c) and (f).
The vertical dashed lines indicate the positions where instabilities occur.
}
\label{fig:RG_1_dd+ex}
\end{figure*}

\section{Role of intervalley exchange interaction}

The degeneracies of $d$/$p$-wave superconductivity and of CDW/SDW susceptibilities are lifted when intervalley exchange interactions $g_{11}$, $g_{31}$, $g_{41}$ are included.
Their bare values depend on microscopic details as we have discussed in Sec.~\ref{sec:hot-spot}.  For example, such interactions can arise from intervalley scattering mediated by optical phonons.  Since the typical phonon frequency is much larger than the mini-band width, intervalley exchange interactions between low-energy electrons may be even attractive.

Figure~\ref{fig:RG_1_dd+ex} shows the RG flows including both density-density and exchange interactions.  With a small $d_{1-}$ [Figs.~\ref{fig:RG_1_dd+ex}(a), (d)], the superconducting instabilities are dominant but a larger $d_{1-}$ favors density-wave states [Figs.~\ref{fig:RG_1_dd+ex}(b), (c), (e), (f)].  Since their initial values are chosen to be small, the change of exchange interactions under RG is considerably smaller than that of the density interactions. Nonetheless,  degeneracies of susceptibilities are lifted by finite exchange interactions.  We show cases for repulsive interaction in Figs.~\ref{fig:RG_1_dd+ex}(a), (b), (d), (e) and for attractive interaction in Figs.~\ref{fig:RG_1_dd+ex}(c), (f).

Roughly speaking, repulsive interaction prefers $d$-SC to $p$-SC and SDW$^-$ to CDW$^-$, and attractive interaction prefers the converse.  This can be seen from our expressions for interaction strengths shown in Eq.~(\ref{eq:interaction_strength_1}), etc. Depending on the choice of intervalley exchange interactions and nesting parameter $d_{1-}$, any of the four orders---$d$-SC, $p$-SC, CDW$^-$, and SDW$^-$---can be the leading instability.

While the bare values of intervalley exchange interactions are hard to obtain accurately, we now discuss another important factor in selecting between $d$- and $p$-wave SC, and between CDW$^-$ and SDW$^-$.
Both SDW and $p$-wave SC (which is spin-triplet) breaks the $\mathrm{SU}(2)$  spin rotational symmetry, while the CDW and $d$-wave SC (which is spin-singlet) do not. With SU(2) symmetry, it is known that thermal fluctuations associated with Goldstone modes prevent any true long-range spin order in two dimensions. This argument suggests that $d$-SC or CDW$^-$ can still be realized at nonzero temperature, even when the leading susceptibility above the ordering temperature is $p$-SC or SDW$^-$.

\section{Electronic Structure of Density-wave states}
\label{sec:density-wave}

We now examine the electronic structure in a CDW$^-$ state and show that at filling the $n=2$, the CDW$^-$ state can be insulating. The same conclusion applies to a collinear SDW$^-$ because it can be mapped to the CDW$^-$ state by performing a sign change on electrons of one spin polarization in one valley.

In density-wave states, the Brillouin zone in momentum space is reduced because of the enlarged unit cell in real space. The previously distinct momentum eigenstates now can hybridize, resulting in a band structure reconstruction. When the number of electrons in the enlarged unit cell is an even integer, the resulting CDW state can be a band insulator.

\begin{figure}
\centering
\includegraphics[width=\hsize]{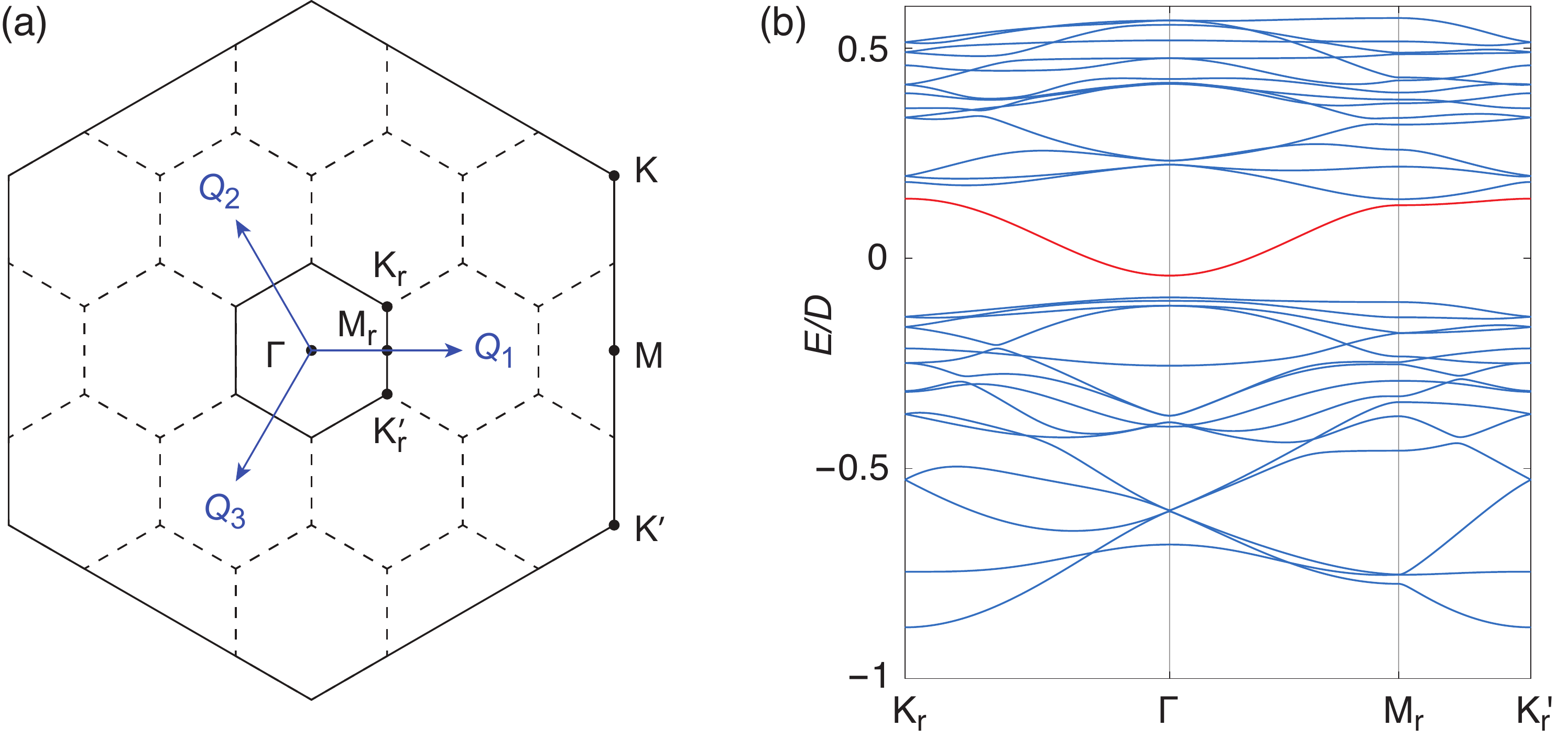}
\caption{
(a) Reduced Brillouin zone in a triple-$Q$ density-wave state.  We assume the three ordering vectors $\bm{Q}_j$ $(j=1,2,3)$, which are parallel to the $\Gamma$--$M$ lines and satisfy $|\bm Q_{j}|=Q^{-}=G/4$.
(b) Energy spectrum in the CDW$^-$ or collinear SDW$^-$ state with order parameter $\Delta =0.16D$. $D$ is the original conduction bandwidth. There are 32 bands in the reduced Brillouin zone and the 17th band from the bottom is colored in red.  Filling of 16 bands corresponds to the filling $n=2$.  
}
\label{fig:CDW-band}
\end{figure}

The CDW$^-$ order can occur at three equivalent wave vectors  related by the $C_3$ rotational symmetry:
\begin{gather}
\begin{gathered}
\bm Q_1 =Q^{-}\bm{n}_0,\quad
\bm Q_2 =Q^{-}\bm{n}_{2\pi/3},\quad
\bm Q_3 =Q^{-}\bm{n}_{-2\pi/3}.
\end{gathered}
\end{gather}
with $\bm{n}_\phi = (\cos\phi, \sin\phi)$. Below we shall consider a triple-$Q$ CDW state, where the above three wave vectors form the new reciprocal vectors, and hence define the reduced Brillouin zone. Compared to a single-$Q$ state, the triple-$Q$ state is expected to be energetically favorable as it gaps more parts of the Fermi surface especially around the hot spots with large DOS.

In the CDW$^-$ state, intervalley order parameter $\langle \psi^\dagger_{\bm{k}+\bm{Q}_i,\tau\sigma} \psi_{\bm{k}\bar{\tau}\sigma} \rangle$ $(\bar{\tau}=-\tau)$ becomes nonzero.
The mean-field Hamiltonian in the CDW$^-$ state thus includes the CDW potential in addition to the original electron dispersion:
\begin{align}
H_{\text{CDW}}
=  \sum_{\bm{k}\tau} \bigg[ \epsilon^\tau_{\bm{k}} \psi^{\dagger}_{\bm k\tau} \psi_{\bm k\tau}
+\Delta \sum_{j=1}^{3}(\psi^{\dagger}_{\bm{k}+\bm{Q}_j,\tau}\psi_{\bm{k}\bar{\tau}}+\mathrm{H.c.}) \bigg].
\end{align}
Since the spin structure is irrelevant, we have dropped the spin index $\sigma$.
Here we assume that the CDW order parameters at $\bm{Q}_1$, $\bm{Q}_2$, $\bm{Q}_3$ are equal, so that the resulting state is invariant under the three-fold rotation.

For the original Fermi surface shown in Fig.~\ref{fig:Fermi_surface}, the CDW$^-$ wave vector connecting a pair of hot spots is close to the commensurate vector ${Q}^- \simeq |{\Gamma\mathrm{M}}|/2 = G/4$, where $G$ is the length of the original reciprocal lattice vectors of twisted bilayer graphene.
(The analytic expression of the energy dispersion in the normal state $\epsilon^\tau_{\bm{k}}$ is given in Appendix~\ref{sec:dispersion}.)
With this choice of CDW wave vector $Q^-$, the reduced Brillouin zone is $4 \times 4$ smaller than the original Brillouin zone and can be constructed as shown in Fig.~\ref{fig:CDW-band}(a).
Since there are two conduction bands (one per valley) in the original Brillouin zone, there are 32 bands in the reduced Brillouin zone.  A complete filling of 16 bands corresponds to the filling of $n=2$, where correlated insulating behavior was experimentally observed.

When the CDW$^-$ order parameter is small, the Fermi surface at the filling $n=2$ is not fully gapped due to imperfect nesting. A full gap appears for $\Delta \gtrsim \Delta_c $. For a realistic Fermi surface with good nesting condition, we find the critical value of the order parameter  $\Delta_{c}=0.15D$, where $D$ is the bandwidth of the original conduction band. The fact $\Delta_c \ll D$ justifies our weak coupling approach.

The gapped energy spectrum in the CDW$^-$ state with $\Delta=0.16D$  is presented in Fig.~\ref{fig:CDW-band}(b). Importantly, we note that the direct gap in the CDW state is located at $\Gamma$ in the reduced Brillouin zone. A single electron pocket (with two-fold spin degeneracy) is present above the gap, while two nearly degenerate hole pockets are present below the gap. The hole pockets have much heavier mass than the electron. These features are consistent with quantum oscillation measurements at densities slightly away from $n=2$, as we shall discuss in the next section.

For the commensurate CDW state with $Q^-=G/4$ considered here, the scattering process labeled by $g_{43}$ carries momentum $2Q'=4Q^-=G$, and thus it is allowed.  This process corresponds to the intervalley exchange interaction and it is presumably smaller than intravalley and intervalley density interactions. We confirm that inclusion of small $g_{43}$ does not alter the RG flow much, and we obtain qualitatively the same result \cite{g43}.

\section{Discussions}

In this section, we compare our results with the experiments on twisted bilayer graphene \cite{exp1}.  We have found from RG analysis the intertwining of unconventional superconductivity and density-wave instabilities. We have obtained from band structure calculations the gapped spectrum of density-wave states at the filling $n=2$.

On the experiment side, the resistivity measurement at zero magnetic field near $n=2$ observes a metallic behavior at high temperatures, then an upturn of resistivity in an intermediate temperature region, before superconductivity appears at the lowest temperature. Furthermore, the in-plane upper critical field of the superconducting state is found to be comparable to the Pauli limit, indicating spin-singlet pairing.  The change from insulating to superconducting behaviors is consistent with the intertwined density wave and SC instabilities, shown by the evolution of susceptibility with decreasing energy scale in Figs.~\ref{fig:phase_dd}(e), (f) and also Figs.~\ref{fig:RG_1_dd+ex}(e)--(f).
Finally, when superconductivity is destroyed by the magnetic field, resistivity becomes insulating at the lowest temperature.

We interpret this $T=0$ insulating state as a CDW/SDW state at wave vector $Q^-$. We have analyzed a triple-$Q^-$ CDW/collinear SDW phase with $4\times 4$ periodicity and have shown that a moderate density-wave order parameter fully gaps the energy spectrum at the filling $n=2$, consistent with the insulating behavior of resistivity at low temperature. Importantly, at densities slightly above $n=2$ (or doping towards complete filling of mini-bands), a single pocket with two-fold degeneracy is found in quantum oscillation measurements. This is consistent with our finding of a single pocket with spin degeneracy above the gap. On the other hand, at densities slightly below $n=2$, quantum oscillations have so far not been observed. This is consistent with the fact that the pockets below the gap in our density-wave state have heavy mass.

\begin{acknowledgments}
We thank Pablo Jarillo-Herrero, Yuan Cao, Valla Fatemi, Oskar Vafek, Leonid Levitov, Sankar Das Sarma, Allan MacDonald, Matthew Yankowitz, Cory Dean and especially Eva Andrei for helpful discussions.
This work is supported by the DOE Office of Basic Energy Sciences, Division of Materials Sciences and Engineering under award DE-SC0010526. LF is partly supported by the David and Lucile Packard Foundation.

\end{acknowledgments}

\appendix

\section{Fermi surface nesting}
\label{sec:nesting}

\subsection{Perfect and near nesting}

Fermi surface nesting provides singularities in susceptibilities.  The susceptibilities (Lindhard functions) in the particle-hole and particle-particle channels are given by
\begin{gather}
\chi^\text{ph}_{\tau\tau'} (\bm{q},\omega)
= \int_{\bm{k}} \frac{f(\epsilon^\tau_{\bm{k}+\bm{q}})-f(\epsilon^{\tau'}_{\bm{k}})}{\omega-\epsilon^\tau_{\bm{k}+\bm{q}}+\epsilon^{\tau'}_{\bm{k}}}, \\
\chi^\text{pp}_{\tau\tau'} (\bm{q},\omega)
= \int_{\bm{k}} \frac{f(\epsilon^\tau_{\bm{k}+\bm{q}})-f(-\epsilon^{\tau'}_{-\bm{k}})}{\omega-\epsilon^\tau_{\bm{k}+\bm{q}}-\epsilon^{\tau'}_{-\bm{k}}}.
\end{gather}
$\tau$ and $\tau'$ denote Fermi surfaces, which correspond to the valley degrees of freedom for the case of twisted bilayer graphene.

The conditions for perfect nesting is given by
\begin{align}
& \epsilon^\tau_{\bm{k}+\bm{q}}=-\epsilon^{\tau'}_{\bm{k}} & &\text{(particle-hole channel)}, \\
& \epsilon^\tau_{\bm{k}+\bm{q}}=\epsilon^{\tau'}_{-\bm{k}} & &\text{(particle-particle channel)}.
\end{align}
When the Fermi surfaces are perfectly nested, i.e., one of the above conditions holds in a certain area of the Brillouin zone, singularities in the susceptibilities are found in the static limit $\omega \to 0$.  One finds a logarithmic divergence in a susceptibility with perfectly-nested Fermi surfaces in two dimensions, so that the susceptibility has $\ln(\Lambda/\omega)$ dependence.

In order to observe logarithmic dependence in susceptibilities at $\omega$, the nesting condition should hold at the energy scale determined by $\omega$.
In other words, if the conditions are approximately met with an accuracy of around $\omega$, we see $\ln(\Lambda/\omega)$ behavior at the energy scale $\omega$.
It allows to relax the nesting conditions at $\omega$ to be
\begin{gather}
\delta^\text{ph}(\omega) \equiv \left| \frac{\epsilon^\tau_{\bm{k}+\bm{q}} + \epsilon^{\tau'}_{\bm{k}}}{\omega} \right| \ll 1, \\
\delta^\text{pp}(\omega) \equiv \left| \frac{\epsilon^\tau_{\bm{k}+\bm{q}} - \epsilon^{\tau'}_{-\bm{k}}}{\omega} \right| \ll 1,
\end{gather}
for the particle-hole and particle-particle channels, respectively.
When Fermi surfaces are perfectly nested, we have $\delta^\text{ph}=0$ or $\delta^\text{pp}=0$, and $\ln(\Lambda/\omega)$ dependence in the susceptibility holds down to the lowest energies.
On the other hand, when Fermi surfaces are nearly nested with $\delta^\text{ph/pp}(\omega)\ll 1$ for $\Lambda_0 < \omega \leq \Lambda$, we see a logarithmic enhancement with in the range $\Lambda_0 < \omega \leq \Lambda$, and it is cut off by the lower bound $\Lambda_0$.

Our RG analysis focus on the energy range $\Lambda_0 < \omega \leq \Lambda$, where Fermi surfaces are regarded as nearly nested, and hence assume the parameters $d_{as}$ are constant within the range.  For energy scale below $\Lambda_0$, $d_{as}$ can no longer be regarded as constant, and we need to consider the $y$ dependence of $d_{as}$.

\subsection{Inner and outer Fermi surfaces}

\begin{figure}
\centering
\includegraphics[width=.8\hsize]{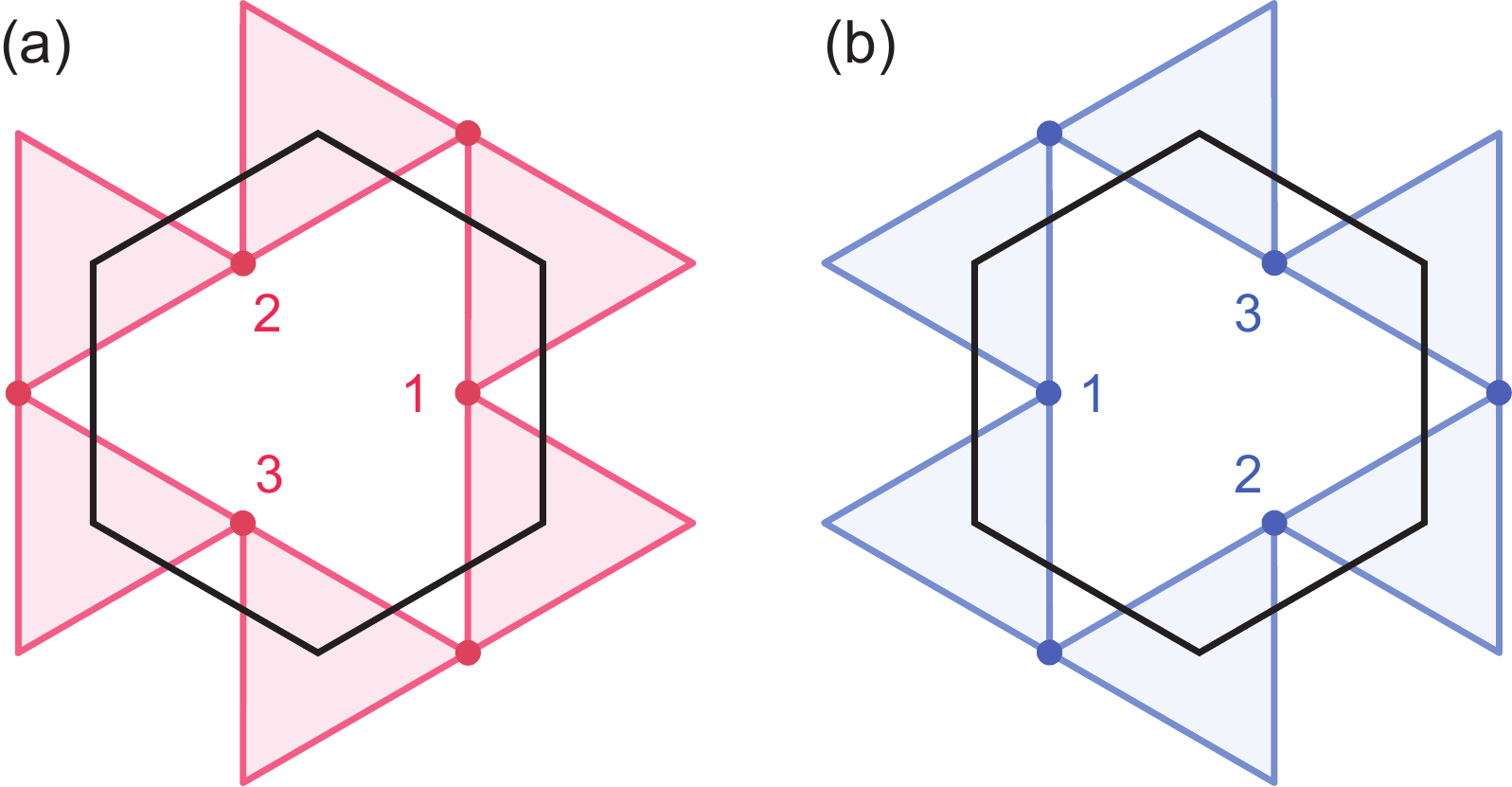}
\caption{
Simplified Fermi surface to show Fermi surface nesting with different wave vectors.  (a) and (b) are Fermi surfaces for different valley degrees of freedom.  Labels $1,2,3$ are the patch indices, assigned at the hot spots.
}
\label{fig:FS_simplified}
\end{figure}

Figure~\ref{fig:FS_simplified} shows simplified Fermi surfaces at the Van Hove energy, where the Fermi surfaces are approximated as corner-sharing triangles, to emphasize Fermi surface nesting.
For convenience, we regard those Fermi surfaces as large hole Fermi surfaces encircling the $\Gamma$ point.
At a saddle point located along the $\Gamma$--$M$ line, a Fermi surface touches another from an adjacent Brillouin zone.
Therefore, there are two Fermi surfaces in one patch around a hot spot when we consider the reduced Brillouin zone.
For convenience, we call the Fermi surface from the first Brillouin zone as the ``inner'' Fermi surface and the other from the second Brillouin zone as the ``outer'' Fermi surface.

From the figure, we find four distinct Fermi surface nestings.
First, the Fermi surfaces of patches $1$ and $1'$ ($1$ and $1'$ are from different valleys) are perfectly nested in the particle-particle channel, yielding the BCS instability.  It always holds for time-reversal-invariant systems.  There are also symmetry-related pairs: $2$--$2'$ and $3$--$3'$.
Second, the inner Fermi surfaces are nested between $1$ and $1'$, leading to the logarithmic dependence in $\chi^\text{ph}(Q',\omega)$ and constant $d_{2-}$.
In addition, by looking at both inner and outer Fermi surfaces, we can find that the Fermi surfaces from $1$ and $2'$ are nested both in the particle-hole and particle-particle channels. It gives logarithmic dependences in $\chi^\text{ph}(Q^-,\omega)$ and $\chi^\text{pp}(Q^+,\omega)$, which results in constant $d_{1-}$ and $d_{3-}$, respectively.

The discussion can be done in parallel near the Van Hove energy.
When the Fermi energy is slightly above the Van Hove energy, we can regard the Fermi surfaces as hole pockets around the $\Gamma$ point similarly as above.
If we consider the filling slightly below the Van Hove energy, we should instead look at electron pockets, which surround the $K$ and $K'$ points; cf.\ Fig.~\ref{fig:Fermi_surface}(c).
In this case, a hot spot involves two Fermi surfaces from the two electron pockets.  Still, a hot spot contains two Fermi surfaces, and Fermi surface nesting for both of them should be considered.

\section{Susceptibilities near the Van Hove energy}
\label{sec:chi_VHS}

The DOS logarithmically diverges at a saddle point in two dimensions because of the Van Hove singularity.  When we approximate the energy dispersion near a saddle point as $\epsilon(\delta\bm{k})=A\delta k_x^2-B\delta k_y^2$, we obtain the DOS (in the vicinity of the saddle point) as
\begin{equation}
\rho_0(\epsilon) = \frac{1}{2\pi^2\sqrt{AB}} \ln\frac{\Lambda}{\epsilon},
\end{equation}
where $\Lambda$ is the high-energy cutoff, which corresponds to the patch size in the present case.

We have defined the eight susceptibilities $\chi_{as}$ in the particle-hole and particle-particle channels with four different wave vectors in Eqs.~\eqref{eq:chi_i}--\eqref{eq:chi_f}.
Among the eight susceptibilities, $\chi_{2+}$ corresponds to the DOS; $\chi_{2+}(\omega) = \rho_0(\omega)$.
The divergence in the limit $\omega \to 0$ is relaxed by finite chemical potential or away from the Van Hove energy:
\begin{equation}
\chi_{2+}(\omega)= \frac{1}{2\pi^2\sqrt{AB}} \ln\frac{\Lambda}{\max(\omega,\mu)}.
\end{equation}

The susceptibility in the BCS channel $\chi_{0-}$ suffers from the Cooper instability in the presence of time-reversal symmetry. An alternative way to state this is that the Fermi surfaces are perfectly nested in this channel.
We note that $\chi_{0-}$ is an intervalley susceptibility since the saddle points at $\bm{k}_{\alpha\tau}$ and $-\bm{k}_{\alpha\tau}(=\bm{k}_{\alpha,\tau'\neq\tau})$ belong to different valleys.
Calculating $\chi_{0-}(\omega)$, we obtain
\begin{equation}
\label{eq:Cooper_channel}
\chi_{0-}(\omega)= \frac{\rho_0(\omega)}{4} \ln\frac{\Lambda}{\omega}
=\frac{1}{8\pi^2\sqrt{AB}}\ln\frac{\Lambda}{\max(\omega,\mu)} \ln\frac{\Lambda}{\omega}.
\end{equation}
For $\mu=0$ (at the Van Hove energy), $\chi_{0-}$ has a double log singularity.  It is the leading divergence in the eight susceptibilities and it sets the RG scale $y$.
It is important to note that the two logarithms have different origins: one logarithmic singularity $[\ln(\Lambda/\max\{\omega,\mu\})]$ is from the Van Hove singularity and the other $[\ln(\Lambda/\omega)]$ from the Fermi surface nesting.
The logarithmic divergence from Fermi surface nesting has already been discussed.  The divergence of the DOS is suppressed away from the Van Hove energy.  However, if $\mu\ll\Lambda$ is satisfied, there still are large spectral weights in the patches around the hot spots, which justifies the use of patch RG.

When Fermi surfaces are nested for a susceptibility in a certain channel and a wave vector (say, $\chi_{as}$), a logarithmic dependence from Fermi surface nesting is present.  (One can find an explicit calculation for the case of monolayer graphene in e.g.\ Ref.~\cite{Gonzalez}.)  Then, $\chi_{as}$ has the same singularity as $\chi_{0-}$ and the corresponding parameter $d_{as}(y)=d\chi_{as}/dy$ becomes constant as we define the RG scale by $y\equiv \chi^\text{pp}(0,\epsilon)$.
In contrast, Fermi surface nesting is absent for $\chi_{as}$, the corresponding parameter $d_{as}(y)$ decays as $y^{-1/2}$ for $y\gg 1$ \cite{Furukawa,Nandkishore}.

\section{Susceptibilities for ordering instabilities}
\label{sec:general_analysis}

\begin{figure*}
\centering
\includegraphics[width=0.6\hsize]{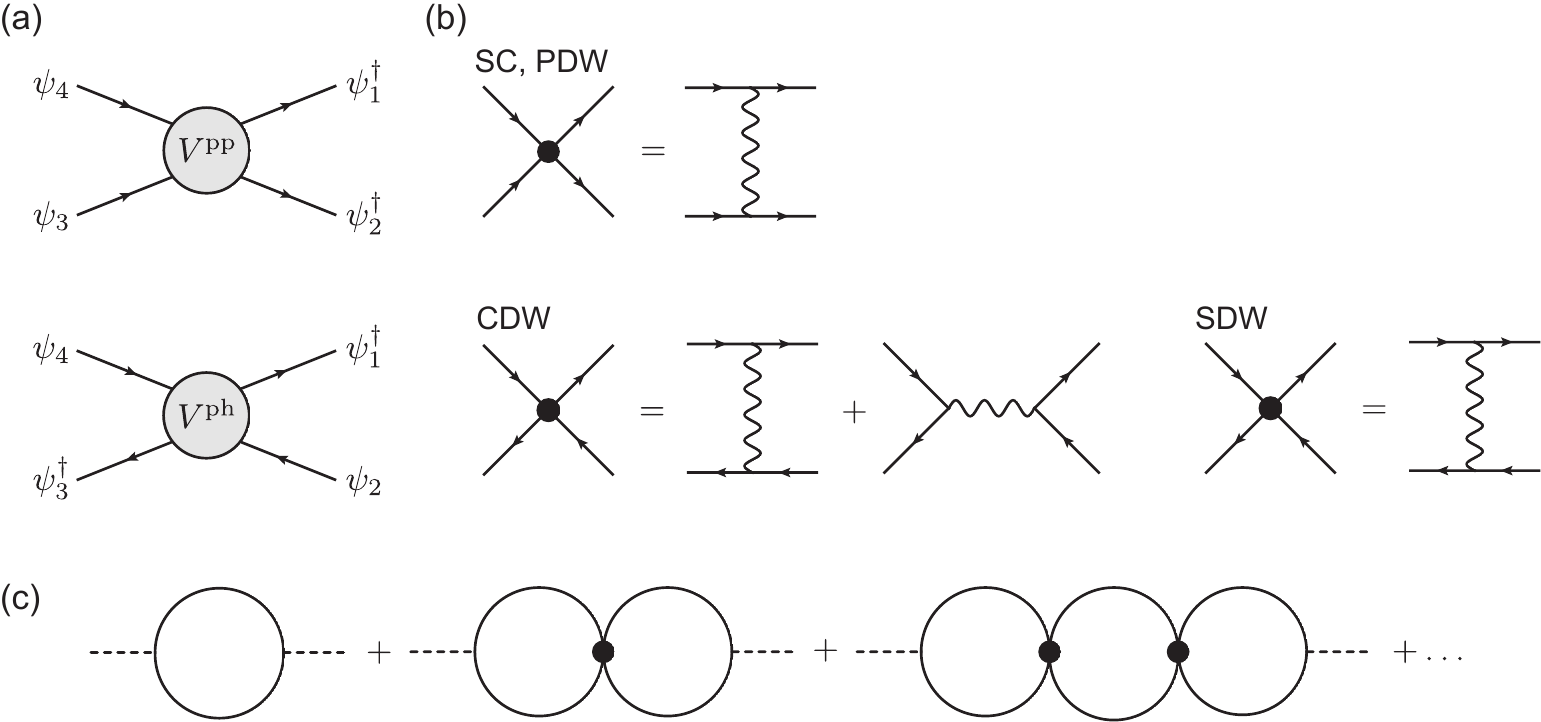}
\caption{
Diagrammatic representation of interactions and summation for the susceptibility.  (a) Diagrams for interactions in the particle-particle channel $V^\text{pp}$ and in the particle-hole channel $V^\text{ph}$. The two-particle interactions are drawn by four-leg ladders.  (b) Interactions for various orderings to the lowest order.  (c) RPA-like resummation for susceptibilities.
}
\label{fig:interaction_susceptibility}
\end{figure*}

\begin{figure*}
\centering
\includegraphics[width=.7\hsize]{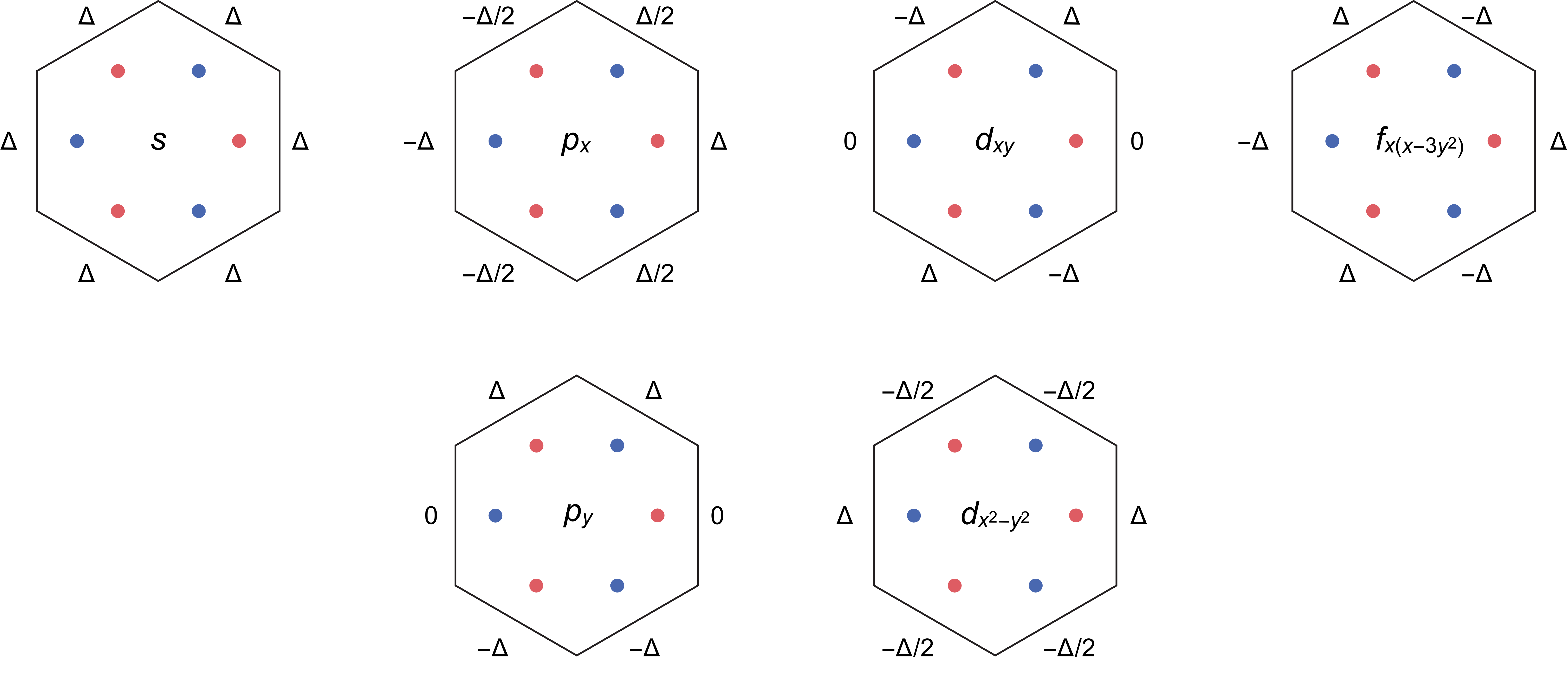}
\caption{
SC pairing symmetries and order parameters at hot spots.  Note that the order parameters are not normalized for each pairing.
}
\label{fig:pairing}
\end{figure*}

\subsection{Interaction strengths for instabilities}

Now we consider an interaction strength that may trigger an ordering instability.  We write two-particle interactions in the particle-particle and particle-hole channels as
\begin{gather}
\hat{V}^\text{ph} = V^\text{ph} \psi_1^\dagger \psi_2 \psi_3^\dagger \psi_4, \\
\hat{V}^\text{pp} = V^\text{pp} \psi_1^\dagger \psi_2^\dagger \psi_3 \psi_4.
\end{gather}
One can perform a mean-field analysis by considering a mean field composed of the first or last two fermion operators on the right-hand sides.   Those two types of interactions are diagrammatically depicted in Fig.~\ref{fig:interaction_susceptibility}(a).

In the present model, there are 16 distinct scattering processes, and to the lowest order, interaction strengths for ordering instabilities are written as linear combinations of those coupling constants [Fig.~\ref{fig:interaction_susceptibility}(b)].
Various interaction strengths $V_\eta$ corresponding to ordering $\eta$ with all 16 coupling constants are given as follows:
\allowdisplaybreaks
\begin{gather}
V_{s\text{-SC}} = 2(g_{42}+g_{41}+g_{32}+g_{31}), \\
V_{p\text{-SC}} = 2(g_{42}-g_{41}-g_{32}+g_{31}), \\
V_{d\text{-SC}} = 2(g_{42}+g_{41}-g_{32}-g_{31}), \\
V_{f\text{-SC}} = 2(g_{42}-g_{41}+g_{32}-g_{31}), \\
\begin{aligned}
V_{\text{CDW}^+} =& 4(g_{12}+g_{14}+g_{32}+g_{34}) \\ &-2(g_{21}+g_{24}+g_{31}+g_{34}),
\end{aligned}\\
\begin{aligned}
V_{\text{CDW}^-} = & 4(g_{11}+g_{13}+g_{31}+g_{33}) \\ & -2(g_{22}+g_{23}+g_{32}+g_{33}),
\end{aligned}\\
\begin{aligned}
V_{\text{CDW}'} = & 2(2g_{41}-g_{42}+g_{43}) \\ & -2(g_{12}+g_{13}) +4(g_{21}+g_{23}),
\end{aligned}\\
V_{\text{SDW}^+} = -2(g_{21}+g_{24}+g_{31}+g_{34}), \\
V_{\text{SDW}^-} = -2(g_{22}+g_{23}+g_{32}+g_{33}), \\
V_{\text{SDW}'} = -2(g_{12}+g_{13}+g_{42}+g_{43}), \\
V_{\text{PDW}^+} = 2(-g_{11}-g_{12}+g_{21}+g_{22}), \\
V_{\text{PDW}^-} = 2(-g_{13}-g_{14}+g_{23}+g_{24}), \\
V_{\text{PDW}'} = 2(g_{33}+g_{34}+g_{43}+g_{44}), \\
V_c = -(g_{11}+g_{14}) +2(g_{22}+g_{24}) -g_{41}+2g_{42}+g_{44}, \\
V_s = -(g_{11}+g_{14}+g_{41}+g_{44}).
\allowdisplaybreaks[0]
\end{gather}
In addition to instabilities for superconductivity and density waves, we also write down the interaction strengths for the charge compressibility $V_c$ and the uniform spin susceptibility $V_s$.

As for superconductivity, we can consider six different pairing symmetries because of the six patches.  The six pairing symmetries are distinct in regard to the superconducting order parameters $\Delta$ at the patches (Fig.~\ref{fig:pairing}).  Each $p$- and $d$-wave pairing has two different pairings, but those two give the same interaction strength $V_\eta$.
We assume that the system respects the point group $D_3$, which has the representations $A_1$, $A_2$, and $E$.  From this viewpoint, $s$- and $f$-wave pairings belong to the one-dimensional representations $A_1$ and $A_2$, respectively, and each $p$- and $d$-wave pairing belong to the two-dimensional representation $E$.
For the other interaction strengths, we assume the $A_1$ representation.

\subsection{Derivation of susceptibilities for ordering instabilities}

We calculate the susceptibilities for ordering $\eta$ $\chi_\eta$ by summing up RPA-like diagrams, shown in Fig.~\ref{fig:interaction_susceptibility}(c); see also Ref.~\cite{Chubukov}.  It yields the susceptibility $\chi_\eta$ in the form of
\begin{align}
\chi_\eta &= \chi_\eta^0 - \chi_\eta^0 V_\eta \chi_\eta^0 + \chi_\eta^0 V_\eta \chi_\eta^0 V_\eta \chi_\eta^0 -\ldots
= \frac{\chi_\eta^0}{1+V_\eta\chi_\eta^0},
\end{align}
where $\chi_\eta^0$ is the susceptibility calculated without interaction.
$\chi_\eta^0$ are equal to $\chi_{as}$ with $\{as\}$ depending on $\eta$:  $0+$ (PDW$'$), $0-$ (all SC), $1+$ (CDW$^+$, SDW$^+$), $1-$ (CDW$^-$, SDW$^-$), $2-$ (CDW$'$, SDW$'$), $3+$ (PDW$^-$), and $3-$ (PDW$^+$).

\section{Analysis of the RG equations without nesting}
\label{sec:RG_analytic}

\begin{figure}
\centering
\includegraphics[width=.6\hsize]{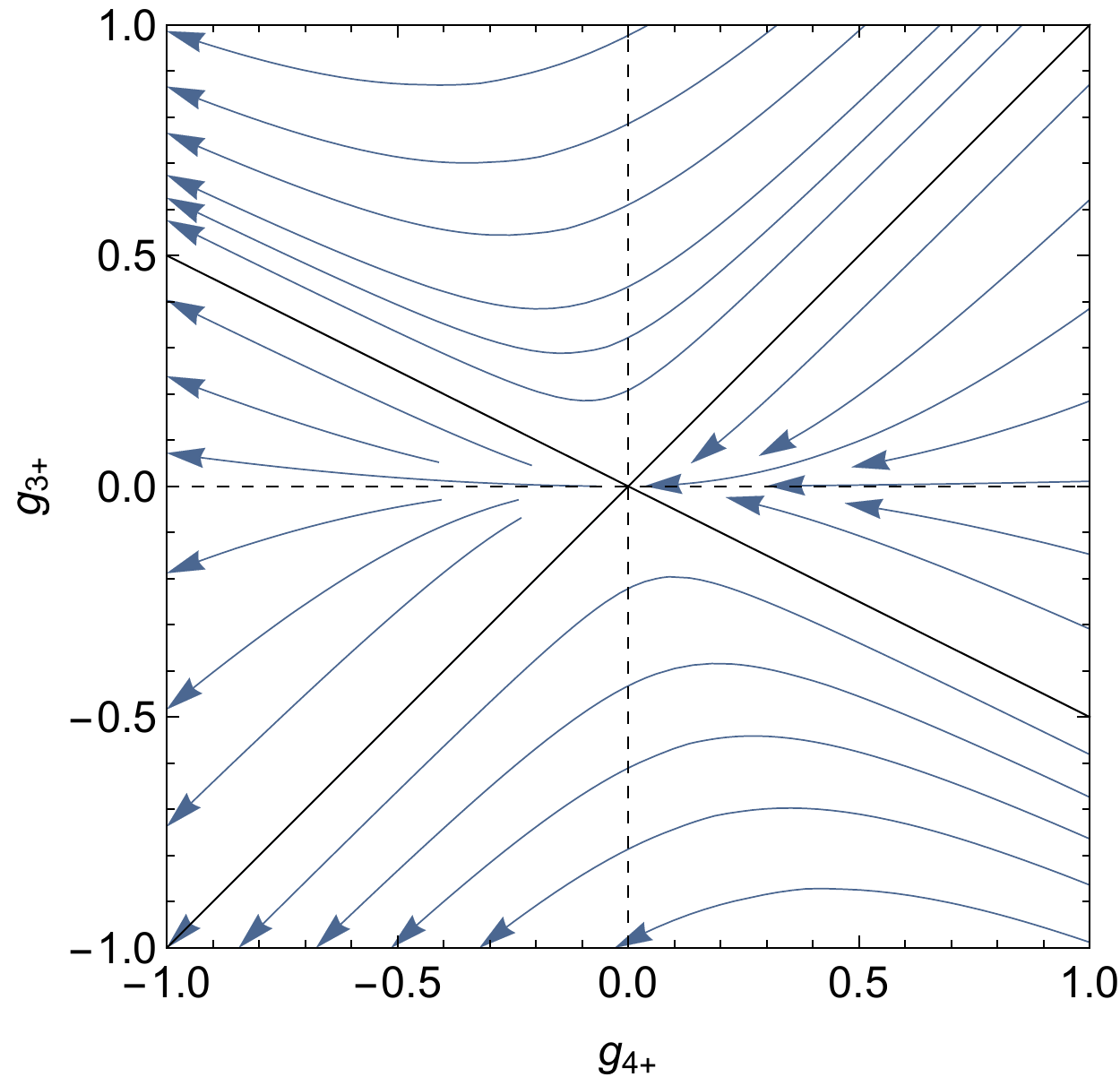}
\caption{
RG flows on the $(g_{4+},g_{3+})$ plane without Fermi surface nesting other than the BCS channel.  Black solid lines are the separatrix lines of the flows.
One finds the same RG flow for $g_{4-}$ and $g_{3-}$.
}
\label{fig:RG_g3_g4}
\end{figure}

The RG equations \eqref{eq:RG_i}--\eqref{eq:rg_f} involve many variables and parameters, and a full analysis of all nine running coupling constants is rather complicated.
Here we analyze the simplest case where there is no Fermi surface nesting other than the BCS channel ($d_{as}=0$, except for $d_{1-}=1$ by definition).
Then, only $g_{31}$, $g_{32}$, $g_{41}$, and $g_{42}$ receive one-loop corrections since they can be regarded as BCS interactions.  The RG equations for the four coupling constants are written as
\begin{gather}
\frac{dg_{3\pm}}{dy} = -g_{3\pm}^2 -2g_{3\pm}g_{4\pm}, \\
\frac{dg_{4\pm}}{dy} = -2g_{3\pm}^2 -g_{4\pm}^2,
\end{gather}
where we define
\begin{equation}
g_{a\pm} = g_{a2} \pm g_{a1} \quad (a=3,4).
\end{equation}
We note that $g_{a+}$ and $g_{a-}$ characterize spin-singlet and spin-triplet SC, respectively.
The RG flow for $g_{4+}$ and $g_{3+}$ is shown in Fig.~\ref{fig:RG_g3_g4}.
The coupled RG equations can be rewritten as
\begin{gather}
\label{eq:RG_43+}
\frac{d(g_{4\pm}+2g_{3\pm})}{dy} = -(g_{4\pm}+2g_{3\pm})^2, \\
\label{eq:RG_43-}
\frac{d(g_{4\pm}-g_{3\pm})}{dy} = -(g_{4\pm}-g_{3\pm})^2,
\end{gather}
and thus are readily solved.

When we neglect the exchange interactions, $g_{4\pm}$ and $g_{3\pm}$ are reduced to be $g_{42}$ and $g_{32}$, respectively.  If we further assume that those two intervalley density-density interactions have the same amplitude $g_{42}=g_{32}$, the equality holds under the RG flow from Eq.~\eqref{eq:RG_43-}.  Each coupling constant obeys the RG equation $\dot{g}=-3g^2$, which is in accordance with Shankar's result for BCS interactions without nesting \cite{Shankar}.

\section{Energy dispersion of a normal state}
\label{sec:dispersion}

\begin{figure}
\centering
\includegraphics[width=\hsize]{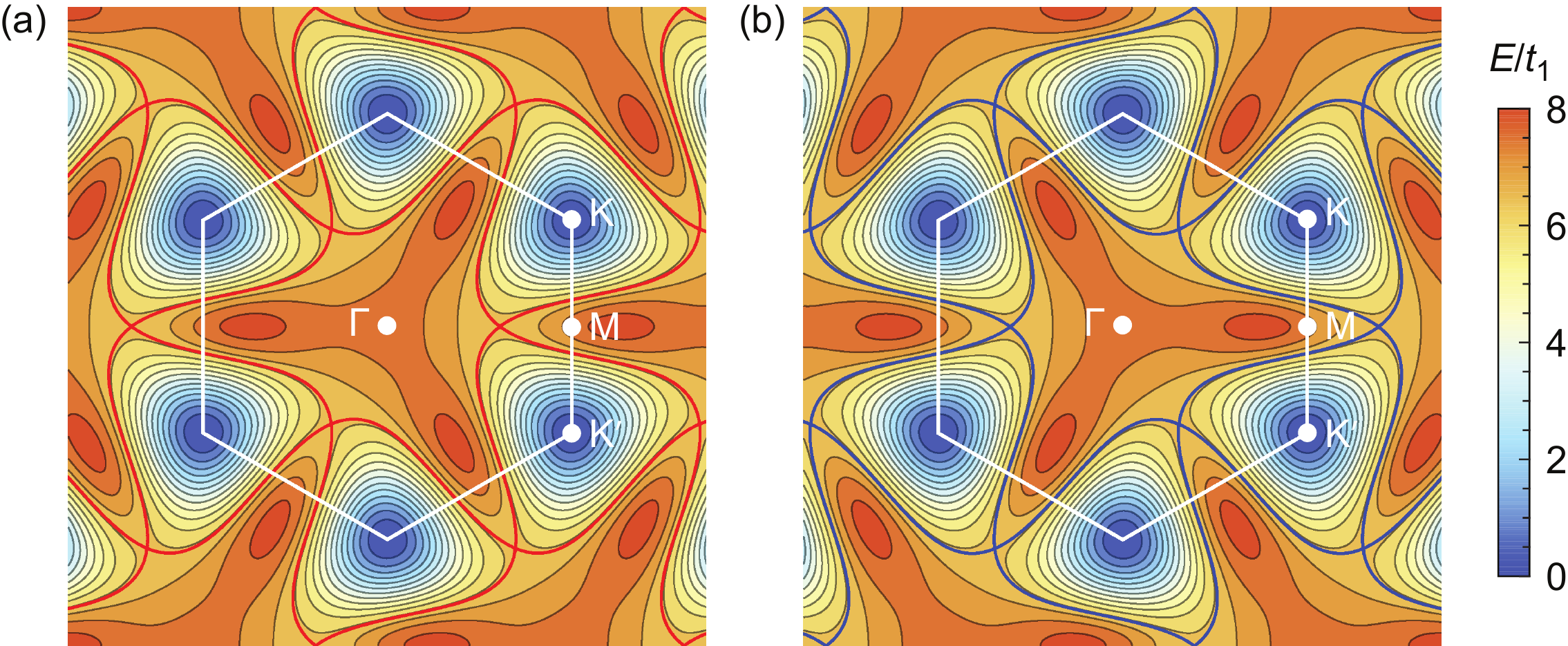}
\caption{
Energy contour plots of the energy bands $\epsilon_{\bm{k}}^\tau$, Eq.~\eqref{eq:energy}.  (a) and (b) correspond to the energy dispersions of the two different valleys.
The parameters are chosen as $t_0=5.4$, $t_1 =1$, $t_2 =-1$, $t'_2 = 0.3$, and $t_3 =0.6$
The Lifshitz transition and accompanying VHS are found at $E/t_1 \approx 6.1$. The bandwidth is $D=7.2$.
}
\label{fig:tight-binding}
\end{figure}

We construct an analytic form of the energy dispersion with two valley degrees of freedom, belonging to the point group $D_3$ and satisfying a filling condition.  For the latter, we require that the energy dispersion have VHS points at the filling $n=2$.
The symmetry conditions are given as follows:
The $C_3$ rotational symmetry is kept within a valley, and hence the energy dispersion satisfies the relation
\begin{equation}
\label{eq:e1}
\epsilon^\tau_{C_3\bm{k}}=\epsilon^\tau_{\bm{k}}.
\end{equation}
There also exist in-plane $C_2$ rotations about the $\Gamma$--$K$ lines, where one of them is taken along the $k_y$ axis.
The in-plane $C_2$ rotation interchanges the two valley and flips $k_x$, thus requiring
\begin{equation}
\label{eq:e2}
\epsilon^\tau_{k_x,k_y}=\epsilon^{\bar{\tau}}_{-k_x,k_y}.
\end{equation}

A function that obeys the symmetry conditions Eqs.~\eqref{eq:e1} and \eqref{eq:e2} can be represented as a Fourier series because of the underlying lattice periodicity.  Up to the third order, the symmetry-allowed terms are given by
\begin{align}
\label{eq:energy}
\epsilon_{\bm{k}}^\tau =
\sum_{i=0}^2 \big[ & t_0 + t_1 \cos(k_y^i) + t_2 \cos(\sqrt{3}k_x^i) + \tau t'_2 \sin(\sqrt{3}k_x^i) \nonumber \\
& + t_3 \cos(2k_y^i) \big],
\end{align}
where we define $k_y^i=C_3^i k_y$ and $k_x^i=C_3^i k_x$.  The parameters $t_1$, $t_2$, $t_2'$, and $t_3$ are real.
In addition to the $D_3$ symmetry, we further impose the filling condition, which is fulfilled by tuning the parameters $t_1$, $t_2$, $t_2'$, and $t_3$.  By choosing e.g.\ $t_1 =1$, $t_2 =-1$, $t'_2 = 0.3$, and $t_3 =0.6$, we find saddle points of the energy dispersion, i.e., VHS points, at the filling $n=2$, as shown in Fig.~\ref{fig:tight-binding}. The energy bands extend in the range of $0 \leq E/t_1 \leq D=7.2$.

\section{Generalized model and RG equations}
\label{sec:generalized_model}

\subsection{Generalized model}

We generalize the model for twisted bilayer graphene, consisting of two valleys with $\mathrm{SU}(2)$ spin, to a $(N_v \times N_s)$-band model, consisting of $N_v$ valleys with $\mathrm{SU}(N_s)$ spins.
We assume that the valley degrees of freedom are nondegenerate in the kinetic part, and as a result, there are $N_v$ separate bands each with $N_s$-fold degeneracy from spin; i.e., the kinetic component has $\mathrm{U}(1)\times\mathrm{SU}(N_s)$ symmetry.
Each Fermi surface is assumed to possess $n_p$ hot spots.  Such a model is applicable for example to cuprates, monolayer graphene, and graphene superlattices. Our results for graphene superlattices are obtained by setting $N_v=N_s=2$ and $n_p=3$.
A model for cuprates corresponds to $N_c=1$, $N_s=2$, and $n_p=2$ \cite{Furukawa,Schulz,Dzyaloshinskii,Lederer,LeHur}, and one for monolayer graphene to $N_c=1$, $N_s=2$, and $n_p=3$ \cite{Nandkishore}.
All those cases are in two dimensions; a generalization to other dimensions is straightforward.

The orbital index takes $\tau = 1,\ldots,N_v$, the spin index $\sigma=1,\ldots,N_s$, and the saddle point index $\alpha=1,\ldots,n_p$.  Redefining the range of the indices, we use the same interaction as Eq.~\eqref{eq:interation}.
As long as we have the same set of coupling constants, the structure of the RG equations does not depend either on the position of the saddle points in the Brillouin zone or the dimensionality of the system, which merely changes the parameters $d_{as}$.  Note that the dimensionality constrains geometrically possible $N_v$ and $n_s$ if we consider the same set of coupling constants.  Also, note that umklapp processes in the valley space $(j=3)$ explicitly violate the valley $\mathrm{U}(1)$ symmetry.

\subsection{RG equations for the coupling constants}

\begin{figure}
\centering
\includegraphics[width=.9\hsize]{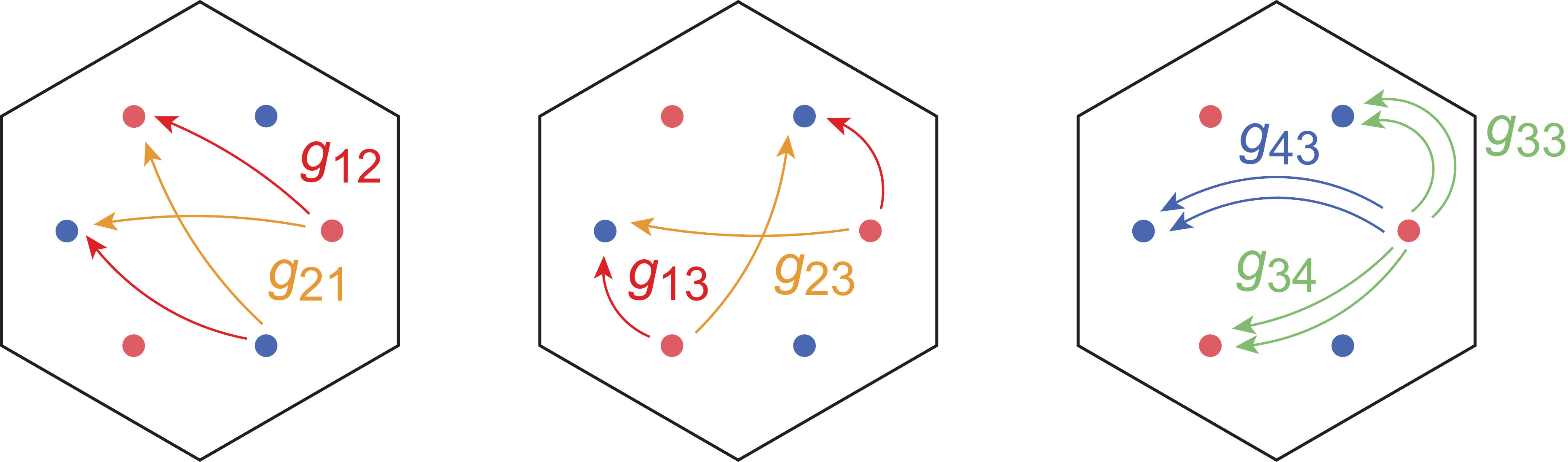}
\caption{
Seven momentum-nonconserving scattering processes.
}
\label{fig:scattering_2}
\end{figure}

For the generalized $(N_v \times N_s)$-band model where each Fermi surface has $n_p$ saddle points, we obtain the RG equations for the sixteen coupling constants $g_{ij}$ to one-loop level (Fig.~\ref{fig:diagram_loop}):
\begin{widetext}
\allowdisplaybreaks
\begin{align}
&\begin{aligned}
\frac{dg_{11}}{dy} =&
    -N_vd_{3-}(g_{11}g_{22}+g_{12}g_{21})
    +N_vd_{2+}(g_{11}g_{44}+g_{14}g_{41}) \\
    &+2d_{1-}[g_{11}g_{22}+g_{31}g_{32}+(N_v-1)(g_{13}g_{23}+g_{33}^2)]
    -N_sd_{1-}[g_{13}^2+g_{33}^2+(N_v-1)(g_{11}^2+g_{31}^2)],
\end{aligned}\\
&\begin{aligned}
\frac{dg_{12}}{dy} =&
    -2d_{3-}[g_{12}g_{22}+(N_v-1)g_{11}g_{21}]
    +2d_{2-}[g_{12}g_{42}+(N_v-1)g_{13}g_{43}] \\
    &+2d_{1+}[g_{12}g_{24}+g_{32}g_{34}+(N_v-1)(g_{14}g_{21}+g_{31}g_{34})]
    -N_vN_sd_{1+}(g_{12}g_{14}+g_{32}g_{34}),
\end{aligned}\\
&\begin{aligned}
\frac{dg_{13}}{dy} =&
    -N_vd_{3+}(g_{13}g_{24}+g_{14}g_{23})
    +N_vd_{2-}(g_{12}g_{43}+g_{13}g_{42}) \\
    &+2d_{1-}[g_{13}g_{22}+g_{32}g_{33}+(N_v-1)(g_{11}g_{23}+g_{31}g_{33})]
    -N_vN_sd_{1-}(g_{11}g_{13}+g_{31}g_{33}),
\end{aligned}\\
&\begin{aligned}
\frac{dg_{14}}{dy} =&
    -2d_{3+}[g_{14}g_{24}+(N_v-1)g_{13}g_{23}]
    +2d_{2+}[g_{14}g_{44}+(N_v-1)g_{11}g_{41}] \\
    &+2d_{1+}[g_{14}g_{24}+g_{34}^2+(N_v-1)(g_{12}g_{21}+g_{31}g_{32})]
    -N_sd_{1+}[g_{14}^2+g_{34}^2+(N_v-1)(g_{12}^2+g_{32}^2)],
\end{aligned}\\
&\begin{aligned}
\frac{dg_{21}}{dy} =&
    -N_vd_{3-}(g_{11}g_{12}+g_{21}g_{22})
    +N_vd_{1+}(g_{21}g_{24}+g_{31}g_{34})
    +d_{2-}[g_{12}g_{41}+g_{21}g_{42}+(N_v-1)(g_{13}g_{43}+g_{23}g_{43})] \\
    &-2N_sd_{2-}[g_{23}g_{43}+(N_v-1)g_{21}g_{41}],
\end{aligned}\\
&\begin{aligned}
\frac{dg_{22}}{dy} =&
    -d_{3-}[g_{12}^2+g_{22}^2+(N_v-1)(g_{11}^2+g_{21}^2)]
    +d_{1-}[g_{22}^2+g_{32}^2+(N_v-1)(g_{23}^2+g_{33}^2)] \\
    &+2d_{2+}[g_{14}g_{42}+g_{22}g_{44}+(N_v-1)(g_{11}g_{44}+g_{24}g_{41})]
    -N_vN_sd_{2+}(g_{22}g_{44}+g_{24}g_{42}),
\end{aligned}\\
&\begin{aligned}
\frac{dg_{23}}{dy} =&
    -N_vd_{3+}(g_{13}g_{14}+g_{23}g_{24})
    +N_vd_{1-}(g_{22}g_{23}+g_{32}g_{33}) \\
    &+2d_{2-}[g_{12}g_{43}+g_{23}g_{42}+(N_v-1)(g_{13}g_{41}+g_{21}g_{43})]
    -N_vN_sd_{2-}(g_{21}g_{43}+g_{23}g_{41}),
\end{aligned}\\
&\begin{aligned}
\frac{dg_{24}}{dy} =&
    -d_{3+}[g_{14}^2+g_{24}^2+(N_v-1)(g_{13}^2+g_{23}^2)]
    +d_{1+}[g_{24}^2+g_{34}^2+(N_v-1)(g_{21}^2+g_{31}^2)] \\
    &+2d_{2+}[g_{14}g_{44}+g_{24}g_{44}+(N_v-1)(g_{11}g_{42}+g_{22}g_{41})]
    -2N_sd_{2+}[g_{24}g_{44}+(N_v-1)g_{22}g_{42}],
\end{aligned}\\
&\begin{aligned}
\frac{dg_{31}}{dy} =&
    -N_v (g_{31}g_{42}+g_{32}g_{41})
      -N_v(n_p-2) g_{31}g_{32}
    +N_vd_{1+}(g_{21}g_{34}+g_{24}g_{31}) \\
    &+2d_{1-}[g_{11}g_{32}+g_{22}g_{31}+(N_v-1)(g_{13}g_{33}+g_{23}g_{33})]
    -2N_sd_{1-}[g_{13}g_{33}+(N_v-1)g_{11}g_{31}],
\end{aligned}\\
&\begin{aligned}
\frac{dg_{32}}{dy} =&
    -2 [g_{32}g_{42}+(N_v-1)g_{31}g_{41}]-(n_p-2) [g_{32}^2+(N_v-1)g_{31}^2]
    +2d_{1-}[g_{22}g_{32}+(N_v-1)g_{23}g_{33}] \\
    &+2d_{1+}[g_{12}g_{34}+g_{24}g_{32}+(N_v-1)(g_{14}g_{31}+g_{21}g_{34})]
    -N_vN_sd_{1+}(g_{12}g_{34}+g_{14}g_{32}),
\end{aligned}\\
&\begin{aligned}
\frac{dg_{33}}{dy} =&
    -N_vd_{0+}(g_{33}g_{44}+g_{34}g_{43})-N_v(n_p-2)d_{0+}g_{33}g_{34}
    +N_vd_{1-}(g_{22}g_{33}+g_{23}g_{32}) \\
    &+2d_{1-}[g_{13}g_{32}+g_{22}g_{33}+(N_v-1)(g_{11}g_{33}+g_{23}g_{31})]
    -N_vN_sd_{1-}(g_{11}g_{33}+g_{13}g_{31}),
\end{aligned}\\
&\begin{aligned}
\frac{dg_{34}}{dy} =&
    -2d_{0+}[g_{34}g_{44}+(N_v-1)g_{33}g_{43}]
      -(n_p-2)d_{0+}[g_{34}^2+(N_v-1)g_{33}^2]
    +2d_{1+}[g_{24}g_{34}+(N_v-1)g_{21}g_{31}] \\
    &+2d_{1+}[g_{14}g_{34}+g_{24}g_{34}+(N_v-1)(g_{12}g_{31}+g_{21}g_{32})]
    -2N_sd_{1+}[g_{14}g_{34}+(N_v-1)g_{12}g_{32}],
\end{aligned}\\
&\begin{aligned}
\frac{dg_{41}}{dy} =&
    -N_v g_{41}g_{42}-N_v(n_p-1) g_{31}g_{32}
    +N_vd_{2+}g_{41}g_{44}+N_v(n_p-1)d_{2+}g_{11}g_{14} \\
    &+2d_{2-}[g_{41}g_{42}+(N_v-1)g_{43}^2]+2(n_p-1)d_{2-}[g_{12}g_{21}+(N_v-1)g_{13}g_{23}] \\
    &-N_sd_{2-}[g_{43}^2+(N_v-1)g_{41}^2]-N_s(n_p-1)d_{2-}[g_{23}^2+(N_v-1)g_{21}^2],
\end{aligned}\\
&\begin{aligned}
\frac{dg_{42}}{dy} =&
    - [g_{42}^2+(N_v-1)g_{41}^2]-(n_p-1) [g_{32}^2+(N_v-1)g_{31}^2] \\
    &+d_{2-}[g_{42}^2+(N_v-1)g_{43}^2]+(n_p-1)d_{2-}[g_{12}^2+(N_v-1)g_{13}^2] \\
    &+2d_{2+}[g_{42}g_{44}+(N_v-1)g_{41}g_{44}]+2(n_p-1)d_{2+}[g_{14}g_{22}+(N_v-1)g_{11}g_{24}] \\
    &-N_vN_sd_{2+}g_{42}g_{44}-N_vN_s(n_p-1)d_{2+}g_{22}g_{24},
\end{aligned}\\
&\begin{aligned}
\frac{dg_{43}}{dy} =&
    -N_vd_{0+}g_{43}g_{44}-N_v(n_p-1)d_{0+}g_{33}g_{34}
    +N_vd_{2-}g_{42}g_{43}+N_v(n_p-1)d_{2-}g_{12}g_{13} \\
    &+2d_{2-}[g_{42}g_{43}+(N_v-1)g_{41}g_{43}]+2(n_p-1)d_{2-}[g_{12}g_{23}+(N_v-1)g_{13}g_{21}] \\
    &-N_vN_sd_{2-}g_{41}g_{43}-N_vN_s(n_p-1)d_{2-}g_{21}g_{23},
\end{aligned}\\
&\begin{aligned}
\frac{dg_{44}}{dy} =&
    -d_{0+}[g_{44}^2+(N_v-1)g_{43}^2]-(n_p-1)d_{0+}[g_{34}^2+(N_v-1)g_{33}^2]
    +d_{2+}[3g_{44}^2+(N_v-1)g_{41}^2+2(N_v-1)g_{41}g_{42}] \\
    &+(n_p-1)d_{2+}[g_{14}^2+2g_{14}g_{24}+(N_v-1)(g_{11}^2+2g_{11}g_{22})] \\
    &-N_sd_{2+}[g_{44}^2+(N_v-1)g_{42}^2]-N_s(n_p-1)d_{2+}[g_{24}^2+(N_v-1)g_{22}^2].
\end{aligned}
\end{align}
% \allowdisplaybreaks[0]
\end{widetext}

The RG equations for the case of twisted bilayer graphene are obtained with $N_v=N_s=2$ and $n_p=3$.
By setting $N_v=1$, $N_s=2$, we can reproduce the previous results for cuprates with $n_p=2$ \cite{Furukawa,Schulz,Dzyaloshinskii,Lederer,LeHur} and for graphene with $n_p=3$ \cite{Nandkishore}.  Since there are no valley degrees of freedom in such cases, we need only four coupling constants $g_i$ and four parameters $d_a$, dropping the second subscripts $j$ from $g_{ij}$ and $s$ from $d_{as}$.


\begin{thebibliography}{99}



\bibitem{exp1}
Y. Cao, V. Fatemi, S. Fang, K. Watanabe, T. Taniguchi, E. Kaxiras, and P. Jarillo-Herrero, Nature \textbf{556}, 43 (2018).

\bibitem{exp2}
Y. Cao, V. Fatemi, A. Demir, S. Fang, S. L. Tomarken, J. Y. Luo, J. D. Sanchez-Yamagishi, K. Watanabe, T. Taniguchi, E. Kaxiras, R. C. Ashoori, and P. Jarillo-Herrero, Nature \textbf{556}, 80 (2018).

\bibitem{Wang}
G. Chen, L. Jiang, S. Wu, B. Lv, H. Li, K. Watanabe, T. Taniguchi, Z. Shi, Y. Zhang, F. Wang, arXiv:1803.01985

\bibitem{Xu}
C. Xu and L. Balents, arXiv:1803.08057.

\bibitem{Volovik}
G. E. Volovik, arXiv:1803.08799.

\bibitem{Yuan}
N. F. Q. Yuan and L. Fu, arXiv:1803.09699.

\bibitem{Po}
H. C. Po, L. Zou, A. Vishwanath, T. Senthil, arXiv:1803.09742.

\bibitem{Zhang}
G.-Y. Zhu, T. Xiang, G.-M. Zhang,  arXiv:1804.00302

\bibitem{Baskaran}
G. Baskaran, arXiv:1804.00627.

\bibitem{Kivelson}
J. F. Dodaro, S. A. Kivelson, Y. Schattner, X.-Q. Sun, C. Wang, arXiv:1804.03162.

\bibitem{Philips}
B. Padhi, C. Setty, P. W. Phillips, arXiv:1804.01101.

\bibitem{Lee}
X. Y. Xu, K. T. Law, and P. A. Lee, arXiv:1805.00478

\bibitem{Yang}
C.-C. Liu, L.-D. Zhang, W.-Q. Chen, and F. Yang, arXiv:1804.10009

\bibitem{TB3}
G. Trambly de Laissardi\`{e}re, D. Mayou and L. Magaud, Nano Lett. \textbf{10}, 804 (2010).

\bibitem{TB2}
E. Su\'{a}rez Morell, J. D. Correa, P. Vargas, M. Pacheco, and Z. Barticevic, Phys. Rev. B \textbf{82}, 121407(R) (2010).

\bibitem{TB1}
P. Moon and M. Koshino, Phys. Rev. B \textbf{85}, 195458 (2012).

\bibitem{TB4}
G. Trambly de Laissardi\`{e}re, D. Mayou, and L. Magaud, Phys. Rev. B \textbf{86}, 125413 (2012).

\bibitem{DFT2}
K. Uchida, S. Furuya, J.-I. Iwata, and A. Oshiyama, Phys. Rev. B \textbf{90}, 155451 (2014).

\bibitem{TB5}
A. O. Sboychakov, A. L. Rakhmanov, A. V. Rozhkov, and F. Nori, Phys. Rev. B \textbf{92}, 075402 (2015).

\bibitem{DFT1}
S. Fang and E. Kaxiras, Phys. Rev. B \textbf{93}, 235153 (2016).

\bibitem{Nori2}
A. V. Rozhkov, A. O. Sboychakov, A. L. Rakhmanov, and F. Nori, Phys. Rev. B \textbf{95}, 045119 (2017).

\bibitem{Nam}
N. N. T. Nam and M. Koshino, Phys. Rev. B \textbf{96}, 075311 (2017).

\bibitem{Neto}
J. M. B. Lopes dos Santos, N. M. R. Peres, and A. H. Castro Neto, Phys. Rev. Lett. \textbf{99}, 256802 (2007).

\bibitem{Mele}
E. J. Mele, Phys. Rev. B \textbf{81}, 161405(R) (2010).

\bibitem{Neto2}
J. M. B. Lopes dos Santos, N. M. R. Peres, and A. H. Castro Neto, Phys. Rev. B \textbf{86}, 155449 (2012).

\bibitem{MacDonald}
R. Bistritzer and A. H. MacDonald, Proc. Natl. Acad. Sci. \textbf{108}, 12233 (2011).

\bibitem{PRL2016}
Y. Cao, J. Y. Luo, V. Fatemi, S. Fang, J. D. Sanchez-Yamagishi, K. Watanabe, T. Taniguchi, E. Kaxiras, and P. Jarillo-Herrero
Phys. Rev. Lett. \textbf{117}, 116804 (2016).

\bibitem{Kim}
Y. Kim, P. Herlinger, P. Moon, M. Koshino, T. Taniguchi, K. Watanabe, and J. H. Smet, Nano Lett. \textbf{16}, 5053 (2016).

\bibitem{Andrei}
G. Li, A. Luican, J. M. B. Lopes dos Santos, A. H. Castro Neto, A. Reina, J. Kong, E.Y. Andrei, Nat. Phys. \textbf{6}, 109 (2010).

\bibitem{Schulz}
H. J. Schulz, Europhys. Lett. \textbf{4}, 609 (1987).

\bibitem{Dzyaloshinskii}
I. E. Dzyaloshinskii, Sov. Phys. JETP \textbf{66}, 848 (1987).

\bibitem{Lederer}
P. Lederer, G. Montambaux, and D. Poilblanc, J. Phys. \textbf{48}, 1613 (1987).

\bibitem{Furukawa}
N. Furukawa, T. M. Rice, and M. Salmhofer, Phys. Rev. Lett. \textbf{81}, 3195 (1998).

\bibitem{LeHur}
K. Le Hur and T. M. Rice, Ann. Phys. \textbf{324}, 1452 (2009).

\bibitem{Nandkishore}
R. Nandkishore, L. S. Levitov, and A. V. Chubukov, Nat. Phys. \textbf{8}, 158 (2012).

\bibitem{1d_1}
N. Menyhard and J. S\'olyom, J. Low Temp. Phys. \textbf{12}, 529 (1973)

\bibitem{1d_1-1}
J. S\'olyom, Adv. Phys. \textbf{28}, 201 (1979).

\bibitem{1d_2}
H. Gutfreund and R. A. Klemm, Phys. Rev. B \textbf{14}, 1073 (1976).

\bibitem{Shankar}
R. Shankar, Rev. Mod. Phys. \textbf{66}, 129 (1994).

\bibitem{Chubukov}
A. V. Chubukov, D. V. Efremov, and I. Eremin, Phys. Rev. B \textbf{78}, 134512 (2008).


\bibitem{Halperin}
B. I. Halperin and T. M. Rice, Rev. Mod. Phys. \textbf{40}, 755 (1968).

\bibitem{g43}
$g_{43}$ is constant when only $d_{1-}$ is finite and it is involved in the susceptibilities for CDW$'$ and SDW$'$, which are irrelevant for the present analysis.  We confirm numerically the qualitatively same result even when both $d_{1-}$ and $d_{2-}$ are nonzero, considering that $g_{43}$ is an intervalley exchange interaction and hence it is likely to be rather smaller than the intravalley interactions and the intervalley density interactions.

\bibitem{Gonzalez}
J. Gonz\'alez, Phys. Rev. B \textbf{78}, 205431 (2008).


\end{thebibliography}
\end{document}